\documentclass[aps,prb,twocolumn,superscriptaddress,showpacs,floatfix]{revtex4}

\usepackage{bm}
\usepackage{amsmath}
\usepackage{amssymb}
\usepackage{graphicx}
\usepackage{color}
\usepackage{soul,xcolor}
\setstcolor{red}

\usepackage{xcolor}\definecolor{ao}{rgb}{0.0,0.0,1.0}

\definecolor{br}{rgb}{1.0, 0.22, 0.0}

\usepackage{epsfig}

\def\nab{{\mbox{\boldmath{$\nabla$}}}}
\newcommand{\up}{\uparrow}
\newcommand{\down}{\downarrow}

\def\m{{\mbox{\boldmath{$\mu$}}}}

\def\sig{{\mbox{\boldmath{$\sigma$}}}}
\input{epsf}
\begin{document}
\title{Rashba spin-splitting of single electrons and Cooper pairs\\
}

\author{R. I. Shekhter}
\affiliation{Department of Physics, University of Gothenburg, SE-412
96 G{\" o}teborg, Sweden}

\author{O. Entin-Wohlman}
\email{oraentin@bgu.ac.il}
\affiliation{Raymond and Beverly Sackler School of Physics and Astronomy, Tel Aviv University, Tel Aviv 69978, Israel}
\affiliation{Physics Department, Ben Gurion University, Beer Sheva 84105, Israel}

\author{M. Jonson}
\affiliation{Department of Physics, University of Gothenburg, SE-412
96 G{\" o}teborg, Sweden}
\affiliation{SUPA, Institute of Photonics
and Quantum Sciences, Heriot-Watt University, Edinburgh, EH14 4AS,
Scotland, UK}

\author{A. Aharony}
\affiliation{Raymond and Beverly Sackler School of Physics and Astronomy, Tel Aviv University, Tel Aviv 69978, Israel}
\affiliation{Physics Department, Ben Gurion University, Beer Sheva 84105, Israel}

\date{\today}

\begin{abstract}
Electric weak links, the term used for those parts of an electrical circuit that provide most of the resistance against the flow of an electrical current,  are important elements of many nanodevices. Quantum dots, nanowires and nano-constrictions that bridge two bulk conductors (or superconductors) are examples of such weak links. Here we consider nanostructures where the electronic spin-orbit interaction is strong in the weak link but is unimportant in the bulk conductors,  and explore theoretically the role of the spin-orbit active weak link (which we call a ``Rashba  spin splitter") as a source of new spin-based functionality in both normal and superconducting devices. Some recently predicted phenomena, including mechanically-controlled spin-  and charge currents  as well as the effect of spin polarization of superconducting Cooper pairs, are reviewed.

\end{abstract}


\maketitle

\section{Introduction} 

\label{Intro}

In classical electrodynamics an {\em electric} field affects the spatial (orbital) motion of a charged particle 
while a {\em magnetic} field also leads to a precession of the magnetic moment of a stationary magnetic particle. 
Additional dynamics occur if the magnetic particle {\em moves} in an {\em electric} field since the spatial motion 
of the particle generates a precession of its magnetic moment. 
This precession occurs because in the reference 
frame of the moving particle the electric field is time-dependent and therefore, according to Maxwell's equations, 
generates a magnetic field. 
It follows that the rate of the electric-field induced precession is proportional to both the momentum {\bf p}
of the particle and to the strength of the electric field {\bf E}, giving rise to a Larmor correction in the kinetic energy of the 
electron of the form  \cite{Al-Jaber.1991,ThomasFactor}
\begin{align}
\label{Uclassical}
\Delta E = \frac{1}{2mc} {\m} \cdot\left({\bf p} \times {\bf E}  \right)\ ,
\end{align}
where $\m$ is the magnetic moment of the electron.
Since electrons carry both charge and magnetic moment (spin),  they are subjected to this type of  coupling between 
orbital and magnetic degrees of freedom,  known as the spin-orbit (SO) interaction. 
Remarkably, if in the classical result (\ref{Uclassical}) one lets $\m\rightarrow (-e/mc)\,{\bf s}$, where
$e=\vert e \vert$ and  ${\bf s}=(\hbar/2)\,\sig$ is the electron spin operator 
[the components of the vector $\sig$   are the Pauli matrices $\sigma_{x,y,z}$], and if one also replaces
$e{\bf E}$ by $\nab V$, 
the gradient of the crystal potential, the result coincides with the SO coupling term ${\cal H}^{}_{\rm so}$ in the Pauli equation (the low-velocity approximation of the Dirac equation), \cite{LandauLifshitz4}
\begin{align}
\label{Uqm}
{\cal H}^{}_{\rm so} =- \frac{\hbar}{4m^2c^2}{\sig}  \cdot [{\bf p}\times \nab V({\bf r}) ]\ .
\end{align}
Being a relativistic effect, the SO coupling is small for free electrons in an external electric field  
but can be quite  large for electrons moving in a crystal. There,   the internal electric (crystal) field 
can be very strong, leading in turn to  spin-split energy bands. This is the case
for crystals lacking spatial inversion symmetry as discovered by Dresselhaus
\cite{Dresselhaus.1955} for zinc blende structures (e.g., GaAs, InSb, and CdTe),  and by Rashba and Sheka \cite{Rashba.1959}
 for wurtzite structures (such as GaN, CdS, and ZnO). 

Other examples of systems without spatial inversion symmetry, which are more relevant in the context of this 
review, are those with a surface or an interface. Motivated by experimental work on semiconductor heterostructures 
at the time, Vas'ko \cite{Vasko.1979} and Bychkov and Rashba \cite{Rashba.1984} 
showed theoretically that a surface potential may induce an SO coupling of the
electrons,  that lifts the spin degeneracy of the energy bands. The main contribution to the crystal field turns out  to be not due to the surface potential itself,  but to its effect on the atomic orbitals near the surface, which become distorted (mixed) so that their contributions to the SO coupling are not averaged 
out by symmetry. Although it is possible in principle to calculate an effective SO Hamiltonian for this case {\em ab initio}, \cite{abinitio}
starting from Eq.~(\ref{Uqm}), or using a semi-quantitative tight-binding approach,  \cite{TB} it is convenient to adopt the phenomenological SO Hamiltonian proposed in 
Ref.~\onlinecite{Rashba.1984}. In the notation of Ref.~\onlinecite{Shekhter.2016b},  this ``Rashba" Hamiltonian, which is valid
for systems with a single high-symmetry axis that lack spatial inversion symmetry, reads 
\begin{align}
{\cal H}^{}_{\rm so} = \frac{\hbar k^{}_{\rm so}}{m^*} \sig \cdot\left( {\bf p} \times \hat{\bf n} \right)\,.
\label{HAM}
\end{align}
Here $\hat{\bf n}$ is a unit vector along 
the symmetry axis (the $c$-axis in a hexagonal wurtzite crystal, 
the growth direction in a semiconductor heterostructure, the direction of an external electric field), 
$m^*$ is the effective mass of the electron, and $k_{\rm so}$ is the strength 
of the SO interaction in units of inverse length, \cite{comment2} usually taken from experiments.

The exploration of the ``Rashba physics" that  follows from Eq.~(\ref{HAM}) is today at the heart of the growing research field of ``spin-orbitronics", a branch of spintronics that focuses on the manipulation of nonequilibrium material properties using the SO coupling (see, e.g., the recent reviews 
Refs.~\onlinecite{Manchon.2015} and \onlinecite{Bihlmayer.2015}). 
 
\begin{figure}
\vspace{-1.5cm} \centerline {\includegraphics[width=10cm]{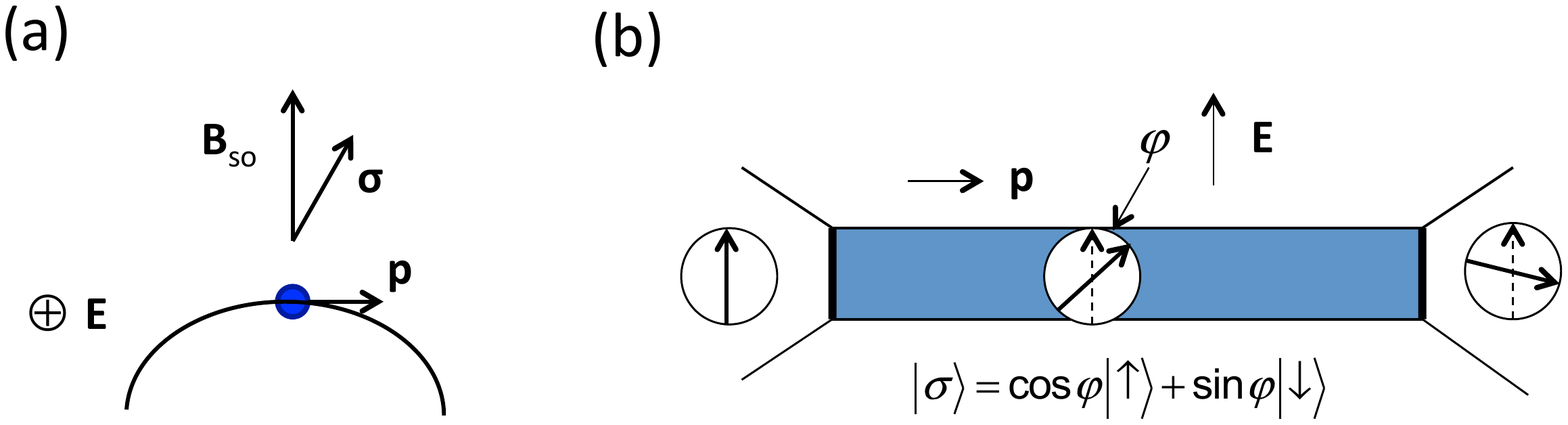}}
\vspace*{-2. cm}
\caption{(a) In a semiclassical picture, the spin ${\bf s} = (\hbar/2)\sig$ [which we for convenience label by $\sig$ rather than by ${\bf s}$]  of an electron moving  with momentum ${\bf p}$ along a curved trajectory
precesses around the effective magnetic field ${\bf B}_{\rm so}$ caused by a spin-orbit (SO) interaction, induced in this simple example
 by an external electric field ${\bf E}$. Being perpendicular to both ${\bf p}$ and ${\bf E}$, the direction of ${\bf B}_{\rm so}$ and hence the direction of the precession axis changes along the trajectory leading to a complex ``spin twisting" effect. (b) The spin evolution for an electron propagating through an SO-active weak link bridging two SO-inactive leads. The spin twist that accompanies the propagation of the electron through the straight one-dimensional wire is pictured as a semiclassical precession of the spin during the time it takes for the electron to pass from the source to the drain electrode.
}
\label{Fig1}
\end{figure}

Semiclassically, the effect of the SO interaction given by Eq.~(\ref{HAM}) on an electron can be viewed as a precession of its spin around an effective magnetic field, ${\bf B}_{\rm so}$, whose direction is perpendicular to both the symmetry axis $\hat{\bf n}$ and the momentum {\bf p} of the electron as it propagates along a trajectory (orbit). When the electron's trajectory is bent, as in Fig.~\ref{Fig1}(a),  the orientation of the precession changes along the trajectory, which makes the picture of the spin evolution more complicated than a simple precession.  We will call the resulting transformation ``spin twisting".

Spin-control of electronic transport can be achieved in principle by incorporating a finite-length SO-active element into a device, for example by using a nanowire made of a material with strong SO coupling as a weak link between two SO-inactive bulk conductors. If  quantum spin-coherence is preserved during the transfer of electrons through the weak link,  the Rashba SO interaction makes it possible to manipulate the spin currents through such devices.
This is because the strong spatial inhomogeneity of the SO coupling prevents the electronic spin from being a good quantum number 
and produces a twisting of the spin of the electrons that enter such a spin-active weak link. 
As we shall see, the net spin twisting accumulated by the electrons as they leave the SO-active weak link  can be controlled by mechanically bending the nanowire and possibly also by using a strong external electric field to tune the SO coupling strength. \cite{Shekhter.2013.2014, Shekhter.2016b}

A semiclassical picture of the spin evolution for an electron propagating through a nanowire-based weak electric link is presented in Fig.~\ref{Fig1}(b) for the simple case of an SO interaction caused by an external electric field. Here the spin twist that accompanies the propagation of an electron through the straight one-dimensional wire is pictured as a semiclassical precession of the spin during the time it takes for the electron to pass through the SO-active wire from the source to the drain electrode.  
Quantum mechanically,  the effect of such a spin rotation can be accounted for by an extra semiclassical phase, $\int \delta {\bf p}\cdot d{\bf r}/\hbar$, which is acquired by the electron wave function because of the renormalization of the electronic momentum, ${\bf p} \to {\bf p} + \delta {\bf p}$, as the electrons enter the weak link. This renormalization is necessary for the total energy to be conserved in the weak link, where the SO interaction modifies the energy.
For a free electron, whose kinetic energy is ${\bf p}^2/(2m)$ before entering the weak link, this extra Aharonov-Casher phase \cite{Aharonov.Casher} follows from Eq.~(\ref{Uclassical}) [where we let $\m \to -e\hbar/(2mc) \sig$]. 
To lowest order in the SO interaction it takes the form 
\begin{align}
\label{ACphase}
\varphi^{}_{\rm AC} = -\frac{e}{4 m c^2} \int (\sig \times {\bf E} ) \cdot d{\bf r} \,.
\end{align}
The Aharonov-Casher phase arises from the interaction between 
the magnetic moment (spin) of an   electron 
and a static electric field. It is dual to the Aharonov-Bohm phase, \cite{Aharonov.Bohm} which is an extra phase induced by the interaction between the
charge of an electron 
and a static magnetic field.

Since $\sig$
is an operator in spinor space,  the Aharonov-Casher phase  (\ref{ACphase}) manifests itself in a splitting of the spin state of an electron that enters the wire (in a ``spin-up" state, say) into a coherent superposition of spin-up and spin-down states. 
We  call such a splitting of electronic waves in spin space ``Rashba spin splitting" and the SO-active weak links that give rise to it ``Rashba (spin) splitters".  

It is instructive to demonstrate the spin splitting explicitly for a simple case. Consider the motion of an electron along the $\hat{\bf x}$ direction. Let ${\bf E} = E \hat {\bf z}$  for $x>0$ and ${\bf E}=0$ for $x<0$,  and $d{\bf r} = dx \, \hat {\bf x}$. 
With the convenient notation ${\tilde k}_{\rm so} = eE/(4 m c^2)$  in Eq.~(\ref{ACphase}), one finds
\begin{align}
e^{i\varphi^{}_{\rm AC}} = e^{-i{\tilde k}_{\rm so}x \,\sigma_y} = \cos({\tilde k}^{}_{\rm so}x) - i \sigma^{}_y \sin({\tilde k}^{}_{\rm so}x).
\end{align}
When this phase factor acts on the spin state $\chi_< = (1, 0)^T = \mid\uparrow \rangle$ (where the subscript $<$ indicates $x<0$), in which the spin is aligned along the positive $\hat{\bf z}$-axis, the resulting spin state is  $\chi_>$ (with $>$ denoting the region $x>0$), where
\begin{align}   
\label{example}
\chi^{}_> =& e^{i\varphi^{}_{\rm AC}}\chi^{}_<  = \left [\begin{array}{c}
\cos({\tilde k}^{}_{\rm so}x) \\
\sin({\tilde k}^{}_{\rm so}x)
\end{array}\right]\nonumber\\
=& \cos({\tilde k}^{}_{\rm so}x) \mid\uparrow\rangle + \sin({\tilde k}^{}_{\rm so}x)\mid\downarrow\rangle\ ,
\end{align}
is a coherent superposition of the spin-up and spin-down states [see also Fig.~\ref{Fig1}(b), where ${\tilde k}^{}_{\rm so}x$ is denoted by $\varphi$]. Note that the explicit form of the initial spin state $\chi_<$ matters for the result of the scattering process described by Eq.~(\ref{example}]. A complementary view comes from noting that $\chi_>$ is an eigenfunction of $\sin({\tilde k}_{\rm so}x) \sigma_x + \cos({\tilde k}_{\rm so}x) \sigma_z$,  which corresponds to a rotation of the spin quantization axis in the $XZ$-plane as the electron propagates along the $\hat{\bf x}$-axis. \cite{comment2}

Before ending this part of the Introduction we mention that strain is another mechanism for inducing an SO coupling, the precise form of which depends on the material and type of strain involved. 
In a single-wall carbon nanotube, for instance, strain can be thought of as occurring when a flat graphene ribbon is rolled up to form a tube. The strain-induced SO coupling in a one-dimensional model of a such a nanotube is described by the Hamiltonian
\begin{align}
{\cal H}^{\rm strain}_{\rm so} = \hbar v_{\rm F }^{}k^{\rm strain}_{\rm so} \sig \cdot \hat{\bf n}\,,
\label{HAMstrain}
\end{align} 
where $v_{\rm F}$ is the Fermi velocity, $k^{\rm strain}_{\rm so}$ is a phenomenological parameter that gives the strength of the SO interaction in units of inverse length, and $\hat{\bf n}$ is a unit vector pointing along the longitudinal axis of the nanotube. Equation (\ref{HAMstrain}) is a simplified form of the SO Hamiltonian derived for a realistic model of such a nanotube. \cite{strain1}

A number of consequences of the Rashba spin splitting for transport phenomena suggested recently will be reviewed here. Two groups of phenomena will be considered. The first concerns incoherent electron transport, where the possible spin-coherence of the Rashba split states does not play any role. In this case the Rashba weak-link can be viewed as a spin-flip scattering center for the transferred electrons. The kinetic consequences of such  a spin-flip relaxation are considered in Sec.~\ref{single} and Sec.  \ref{spin-selective}. Spin-coherent effects in non-superconducting devices only occur in multiply-connected geometries \cite{Ora.and.Amnon} and are outside the scope of this review. Non-trivial interference effects in singly-connected geometries do occur in superconducting structures; these are reviewed in Sec.~\ref{Josephson}.
In particular,  the way by which a supercurrent \cite{Josephson} flowing through a weak link acting as a Rashba spin splitter is affected by the SO interaction is rather unique. 
An   SO-active superconducting weak link brings the opportunity to affect the spin-sensitive pairing of electrons in the superconducting condensate and can be a tool that allows a spin design of superconducting Cooper pairs. Some immediate consequences of such a spin-polarization of the Cooper pairs are presented in  Sec. \ref{Josephson}.

Is the Rashba spin-splitting experimentally important? Clearly, for this to be the case it is necessary that $\varphi_{\rm AC}\sim 1$; the Aharonov-Casher phase accumulated during the propagation of an electron through the spin-split device under consideration must be of order one. In a free-electron model $\varphi_{\rm AC}$ is given by Eq.~(\ref{ACphase}). Its magnitude for a straight SO-active channel of length $d$, placed in a perpendicular electric field $E$,  and  when the  spin-polarization axis is roughly perpendicular to both the electric field and the direction of the channel, is  $\varphi_{\rm AC}  \sim (eEd)/(4mc^2)$.
Since $mc^2=0.5\,$MeV, this gives  for $d=1 \,\mu$m a rather small value,  $\varphi_{\rm AC} \sim 10^{-3}$,  even for an electric field as strong as $E=1\,$V/nm. Allowing for an effective electron mass $m^*\ne m$ and a $g$-factor different from two would add a factor $(gm/2m^*)$ which could be significant if the effective mass is small and the $g$-factor is  large. Even so it seems challenging --- although perhaps not impossible --- to find a situation where the Rashba spin-splitting directly due to an external electric field is important.   An external electric field may well have an indirect effect on the SO interaction, by influencing the mixing of the atomic orbitals particularly in nanoscale systems with poor screening and large surface to volume ratios.

To estimate the scale of the SO interaction due to crystal fields associated with atomic orbitals (in a crystal lacking spatial inversion symmetry) one may consider the electric field at a small distance $r$ from an atomic nucleus of charge $Ze$, given in SI units by the expression $E=Ze/(4\pi\epsilon_0 r^2)$ where $\epsilon_0=8.9\times 10^{-12}$ F/m is the permittivity of vacuum. For $Z=10$ and $r=0.05\,$nm ($=a_{\rm B}$, the Bohr radius) one finds $E \sim 5\times 10^{12}\,$V/m, which in conjunction with Eq.~(\ref{ACphase}) gives $\varphi_{\rm AC} \sim 1$ for the same 1~$\mu$m long channel as above. Rather than trying to improve on this estimate by a full band-structure calculation,  it is common to determine the SO coupling strength from experiments.
Finally, we note that when the SO interaction in a nanowire bridging two bulk (SO-inactive)  electrodes is induced by a crystal field, then although the direction of the crystal field cannot be independently controlled, the spin precession axis in the wire can still be varied with respect to the spin quantization axes in the bulk electrodes by bending the wire.
Generally, a large mechanical deformability of nanostructures, originating from their composite nature complemented by the strong Coulomb forces accompanying single-electron charge transfer, offer an additional functionality of electronic nanodevices. \cite{Hong,Blencowe}  Coherent nano-vibrations in suspended nanostructures, with frequency in the gigahertz range, were detected experimentally. \cite{OConnell}

\section{Suspended nanowires as mechanically-controlled spin splitters }

\label{single}

In charge transport, electronic beam-splitters (e.g., using tunnel barriers) are key ingredients in interference-based devices. Tunnel-barrier scatterers may serve as coherent splitters of the electronic spin when the tunneling electrons also undergo spin (Rashba) scattering. This allows one to map various interference-based phenomena in charge transport onto electronic spin transportation. Such spin-splitters can be  made functional by adding to them a mechanical degree of freedom that controls their geometrical configuration in space, to which the Rashba interaction is quite sensitive. Because of this, one achieves mechanical coherent control and mechanical tuning of the spin filters. \cite{COMsingle} 

\begin{figure}[htp]
\includegraphics[width=5cm]{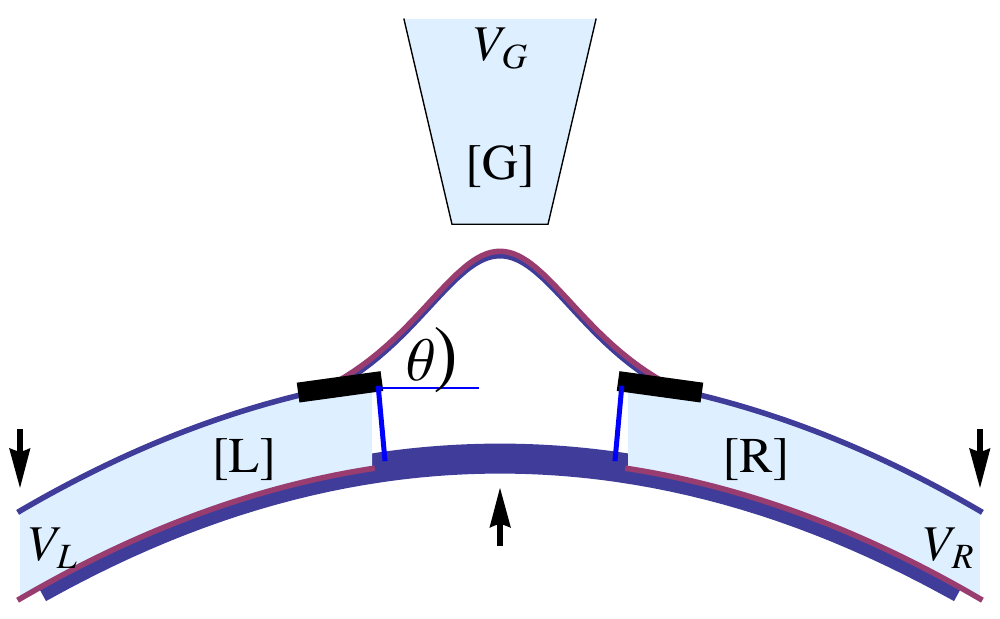}
\caption{A break junction supporting  a nanowire of length $d$  
attached by tunnel contacts  to
two biased electrodes ([L] and [R]). The small vibrations of the wire induce oscillations in the angle $\theta$ 
around some value $\theta_{0}$. The upper electrode ([G]) is an STM
tip biased differently. 
The Rashba interaction can be controlled via the bending angle $\theta$ of the wire. The latter  can be modified both mechanically,
by loads (shown by the arrows) applied to the substrate and electrically, by biasing the STM. 
Reprinted figure with permission from R. I.  Shekhter {\em et al.}, Phys. Rev. Lett. {\bf 111}, 176602 (2013) \textcopyright 2013 by the American Physical Society.
}
\label{Fig3}
\end{figure}

A suspended nanowire, acting as a weak link between two electronic reservoirs, is a good candidate for such a Rashba spin-splitter. \cite{Shekhter.2013.2014}
The amount of
spin splitting brought about by the Rashba interaction on the weak link can be controlled by bending the wire. This can be mechanically tuned, by exploiting a break junction as a substrate for the wire (see Fig.~\ref{Fig3}) or by electrically inducing a Coulomb interaction between the wire and an STM tip electrode (also displayed in Fig.~\ref{Fig3}). This Rashba scatterer is localized on the nanowire, and serves as a pointlike scatterer in momentum-spin space for the electrons incident from the bulky leads. When there is a spin imbalance population in one of the leads (or both), and the Rashba spin-splitter is activated (i.e., the weak link is open for electronic propagation) spin currents are generated and are injected from the pointlike scatterer into the leads. The Rashba splitter thus redistributes the spin populations between the leads. This source of spin currents need not be accompanied by transfer of electronic charges.
We emphasize that although this setup  is similar in the latter aspect to the Datta-Das one, \cite{Das}  our splitter is functional even when the leads are unbiased.

Such a coherent scatterer, whose scattering matrix can be ``designed"   at will by tuning controllably the geometry, can be realized in electric weak links based on clean carbon nanotubes (CNT). Carbon nanotubes have a significant Rashba SO coupling (mainly due to the strain associated with the tube curvature). \cite{Jhang,Kuemmeth,Huertas} 
Moreover, CNT's  are known to have quite long mean-free paths (longer for suspended tubes than for straight ones), allowing for experimental detections of interference-based phenomena (e.g., Fabry-Perot interference patterns). \cite{Biercuk} 
Further tunability of the Rashba spin-splitter can be achieved by switching on an external magnetic field, coupled to the wire through the Aharonov-Bohm effect. \cite{Aharonov.Bohm} This is accomplished by quantum-coherent displacements of the wire, which generate a temperature dependence in the Aharonov-Bohm magnetic flux (through an effective area). \cite{Glazman}

The model system exploited in the calculations is depicted in Fig.~\ref{Fig4}. The tunneling amplitudes through the weak link are calculated in Appendix \ref{app.tun}. It is shown there 
that the linear Rashba interaction manifests itself as a matrix phase factor on the tunneling amplitude. \cite{Gefen} In the geometry of Fig.~\ref{Fig4}, this phase is induced by an electric field perpendicular to the $XY$ plane [see Eq.~(\ref{HAM})], with ${\bf R}_{L}=\{x_{L},y_{L}\}$
for the left tunnel coupling and 
${\bf R}_{R}=\{x_{R},-y_{R}\}$
for the right one, where both radius vectors ${\bf R}_{L}$ and ${\bf R}_{R}$ are functions of the vibrational degrees of freedom (as specified in Sec. \ref{pol}). The quantum vibrations of the wire which modify the bending angle, make the electronic motion effectively two dimensional. This leads to the possibility of manipulating the junction via the Aharonov-Bohm effect, by applying a magnetic field which imposes a further phase on the tunneling amplitudes 
$\phi_{L(R)}=-(\pi/\Phi_{0})(Bx_{L(R)}y_{L(R)})$, where $\Phi_{0}$ is the flux quantum (a factor of order one is absorbed \cite{Glazman} in $B$).

\begin{figure}[htp]
\includegraphics[width=7cm]{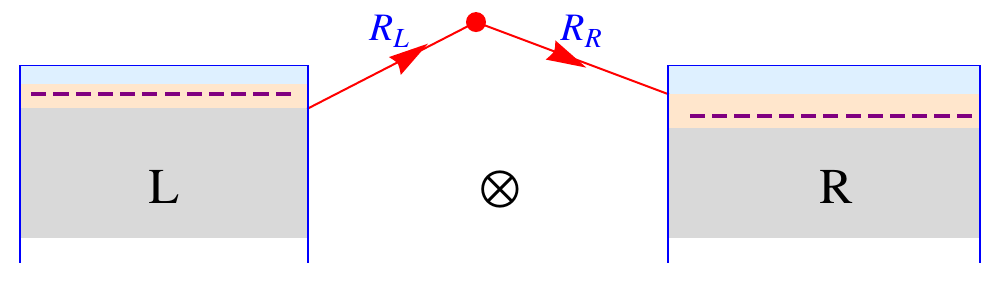}
\caption{
Illustration of the geometry used to calculate the spin-orbit coupling dependence of the tunneling amplitude. Two straight segments are tunnel-coupled to left L and right R electronic electrodes, with possibly different, spin-dependent,  chemical potentials $\mu_{L,\sigma}$ and $\mu_{R,\sigma}$. The setup lies in the $XY$ plane; a magnetic field applied along $\hat{\bf z}$ is shown by $\otimes$.
The setup corresponds to a configuration in which the wire is controlled only mechanically, and the STM is not shown. 
Reprinted figure with permission from R. I.  Shekhter {\em et al.}, Phys. Rev. Lett. {\bf 111}, 176602 (2013) \textcopyright 2013 by the American Physical Society.
}
\label{Fig4}
\end{figure}

The  calculation of the spin-resolved current  through such a junction is detailed in Appendix ~\ref{app.src}, see in particular Sec.~\ref{vib}, Eq.~(\ref{SSD}). 
The flux of electrons  of spin $\sigma$ emerging from the left terminal can be presented in the form
\begin{align}
&I^{}_{L,\sigma}=2 {\cal N}^{}_{L}{\cal N}^{}_{R} \sum_{\sigma '}\sum_{n,n'=0}^{\infty}P(n)
{\cal T}^{}_{nn',\sigma \sigma'}
\nonumber\\
&\times
\Big (1-e^{\beta(\mu^{}_{L,\sigma}-\mu^{}_{R,\sigma'})}\Big )\frac{\mu^{}_{L,\sigma}-\mu^{}_{R,\sigma '}+(n-n')\omega^{}_{0}}{e^{\beta[\mu^{}_{L,\sigma}-\mu^{}_{R,\sigma '}+(n'-n)\omega^{}_{0}]}-1} \ , 
\label{SR}
\end{align}
where ${\cal N}_{L(R)}$ is the density of states at the common chemical potential of the left (right)  lead, and ${\cal T}$ is the spin-dependent  transmission    \cite{com2}
\begin{align}
{\cal T}^{}_{nn',\sigma \sigma'}=|W^{}_{0}|^{2}|\langle n|[e^{-i\psi^{}_{R}}\times e^{-i\psi^{}_{L}}]^{}_{\sigma ',\sigma}|n'\rangle|^{2}\ .
\label{Tt}
\end{align}
Here,  $W_{0}$ is the transmission amplitude in the absence of the SO interaction. (The configuration in which the densities of states are spin dependent is discussed in Sec. \ref{spin-selective}.) In Eq.~(\ref{SR}),  the free vibrations are described by the Einstein Hamiltonian (\ref{ein}) with frequency $\omega_{0}$, $n$ is the vibrationsÕ quantum number, and the weight function $P(n)$ is given in Eq.~(\ref{P}).
For the geometry of Fig.~\ref{Fig4}, 
\begin{align}
\psi^{}_{L}&=\phi^{}_{L}-k^{}_{\rm so}(x^{}_{L}\sigma^{}_{y}-y^{}_{L}\sigma^{}_{x})\ ,\nonumber\\
\psi^{}_{R}&=\phi^{}_{R}-k^{}_{\rm so}(x^{}_{R}\sigma^{}_{y}+y^{}_{R}\sigma^{}_{x})\ .
\label{plr}
\end{align}
The flux of particles 
emerging from the right lead is obtained upon interchanging the roles of the left and the right side of the junction in Eq.~(\ref{SR}).
One notes \cite{Shekhter.2013.2014}  that while the phase due to the magnetic field disappears in the absence of the vibrations, this is not so for the spin-orbit-phase  \cite{Ora.and.Amnon} (as $\psi_{L}$ and $\psi_{R}$ do not commute).

Combining the expressions for the incoming spin currents [Eq.~(\ref{SR}) and the corresponding one for $I_{R,\sigma}$] yields a net spin current, which is injected from the Rashba scatterer into the leads. Therefore, the scatterer can be viewed
as a source of spin current maintained when the leads have imbalanced populations. The spin current, 
\begin{align}
J^{}_{\rm spin}=\sum_{\sigma}J^{}_{{\rm spin},\sigma}=\sum_{\sigma}(I_{L,\sigma}+I^{}_{R,\sigma})\ ,
\end{align} 
tends to diminish the spin imbalance in the leads, through spin-flip transitions induced by the Rashba interaction.
In the limit of weak tunneling, we expect the spin imbalance to be kept constant in time by injecting
spin-polarized electrons into the reservoirs, so that the (spin-dependent) chemical potentials do not vary.

The explicit expressions for the two spin currents yield dramatic consequences. (i) Independent of the choice of the spin-quantization axis, $J_{{\rm spin},\sigma}$ is given solely by the term with 
$\sigma '=\overline{\sigma}$ in Eq.~(\ref{SR}) and the corresponding one for $I_{R,\sigma}$ ($\overline{\sigma}$
is the spin projection opposite to $\sigma$). This implies that only the off diagonal amplitudes  (in spin space) contribute.
(ii) Adopting the plausible geometry detailed in Sec. \ref{pol} [see the discussion preceding Eq.~(\ref{tet})]
one finds
\begin{align}
&e^{-i\psi^{}_{R}}e^{-i\psi^{}_{L}}=e^{i\frac{\pi B d^{2}}{4\Phi^{}_{0}}\sin (2\theta)}
(1-2\cos^{2}\theta\sin (k^{}_{\rm so}d/2)
\nonumber\\
&+i\sigma^{}_{y}\cos\theta\sin (k^{}_{\rm so}d)-i\sigma^{}_{z}\sin (2\theta)\sin^{2}(k^{}_{\rm so}d/2)\ .
\label{pro}
\end{align}
This result is independent of the choice of the spin polarizations in the leads, and does not involve $\sigma_{x}$. (iii) As
Eq.~(\ref{pro}) indicates, spin flips are realized  for any orientation of the leads' polarization. Furthermore, 
when the average angle $\theta_{0}$ [see Eq.~(\ref{tet})] differs from zero,  then both terms on the second line in Eq.~(\ref{pro}) yield spin flips even for the non-vibrating wire. In this respect, the spin-orbit splitting effect is very different from that of the Aharonov-Bohm phase. As mentioned,  the latter  requires the transport electrons to cover a  finite 
area and therefore in our setup is entirely caused by the mechanical vibrations. When $\theta_{0}$ vanishes, 
there are spin flips only if the polarization is in the $XZ$ plane. To be 
concrete, we present below explicit results for a quantization axis along $\hat{\bf z}$. The more general configuration is considered in Sec. \ref{spin-selective}.

In the linear-response regime the spin current loses its dependence on the bias voltage (expressions for these currents beyond 
linear response  are given in  Ref.~\onlinecite{Shekhter.2013.2014})
and becomes
\begin{align}
J^{}_{{\rm spin},\up}=-UG^{}_{\rm spin}\ ,
\end{align}
where
it has been used that
\begin{align}
\mu^{}_{L(R),\up}=\mu^{}_{L(R)}+\frac{U^{}_{L(R)}}{2}
\ ,\ \ 
\mu^{}_{L(R),\down}=\mu^{}_{L(R)}-\frac{U^{}_{L(R)}}{2}
\ ,
\label{mlr}
\end{align}
such that $U=(U_{L}+U_{R})/2$, 
and the spin conductance $G_{\rm spin}$ is
\begin{align}
&G^{}_{\rm spin}=G^{}_{0}\sin^{2}(k^{}_{\rm so}d)\sum_{n=0}^{\infty}\sum_{\ell =1}^{\infty}P(n)\nonumber\\
&\times|\langle n| e^{
i\frac{\pi Bd^{2}}{4\Phi^{}_{0}}\sin(2\theta)}\cos\theta|n+\ell\rangle |^{2}\frac{2\ell\beta\omega^{}_{0}}{e^{\beta\ell\omega^{}_{0}}-1}\ .
\label{GSP}
\end{align}
Here $G_{0}$ is the zero-field electrical conductance divided by $e^{2}$, and $\beta=1/(k_{\rm B}T)$ is the inverse temperature. 
The amount of spin intensity is obtained upon expanding $\theta$  in the operators of the vibrations [see Eq.~(\ref{tet})]. One then finds
\begin{align}
&G^{}_{\rm spin}
=\sin^{2}(k^{}_{\rm so}d)\cos^{2}\theta^{}_{0}G(B)\ ,
\label{G(B)}
\end{align}
where $G(B)$ is precisely the magnetoresistance (divided by $e^{2}$) of the wire, as analyzed in Ref.~\onlinecite{Glazman}, 
and thus has the same behavior at low and high temperatures (as compared to the vibrations' $\omega_{0}$). 
In particular,
\begin{align}
G(B)=|W^{}_{0}|^{2}\Bigg \{\begin{array}{c}1-\frac{\beta\omega^{}_{0}}{6}\frac{B^{2}}{B^{2}_{0}}\ ,\ \ \ \ \ \ \ \ \ 
\beta\omega^{}_{0}\ll 1\ ,\\
\exp[-B^{2}_{}/B^{2}_{0}]\ ,\ \ \ \  \beta\omega^{}_{0}\gg 1\ ,\end{array}\ ,
\label{LIMG}
\end{align}
where 
$B^{}_{0}=\sqrt{2}\Phi^{}_{0}/[\pi da^{}_{0}\cos (\theta^{}_{0})\cos (2\theta^{}_{0})]$ ($a_{0}$ is the amplitude of the zero-point oscillations and $\Phi_{0}$ is the flux quantum).

\section{Spin-resolved transport}
\label{spin-selective}

The formalism presented   in Appendix \ref{app.src}  enables us to study the case where the current through a mechanically-deformed weak link is provided by a battery of uncompensated electronic spins.
When the magnetic polarizations in the electronic reservoirs forming the electrodes are not identical,  then quite generally  both charge and spin currents result from the  transport of electrons through the junction. 
The situation at hand  resembles in a way thermoelectric transport in a two-terminal junction: 
the two currents (charge and spin), flow in response to two affinities, the voltage difference and the difference in the amount of magnetic polarization between  the two reservoirs.
``Non-diagonal"
phenomena, analogous to  the thermoelectric Seebeck and Peltier effects, can therefore be expected. For instance, it is possible to generate  a spin current by  injecting charges  into the material,  
which in turn may give rise to  a spatially inhomogeneous 
spin accumulation.

However, 
the two opposite spins can still contribute   equally   to the charge transport, 
resulting in zero net spin propagation, much like the vanishing of the thermopower when electron-hole symmetry is maintained. In the case of  combined spin and charge transport,   non-diagonal  spin-electric effects appear once the spin and charge transports are coupled in 
a way that distinguishes between the two spin projections.
One may achieve such a spin-dependent transport by exploiting  magnetic materials in which the electronic energy is spin-split. When the magnetization is spatially inhomogeneous (as happens in composite magnetic structures)  the spin-dependent part of the energy is inhomogeneous as well, leading to a spin-dependent force acting on the charge carriers. 
Another possibility, feasible   even in magnetically-homogeneous materials,  is to exploit the     Rashba       SO interaction.  
When this interaction   varies in space, the electronic spin is twisted. 
The end result is the same as in the first scenario above: a spin-dependent force (resulting from the Rashba interaction) 
is exerted on the electrons,  opening the way for non-diagonal  spintro-electric transport.

\begin{figure}[htp]
\includegraphics[width=8cm]{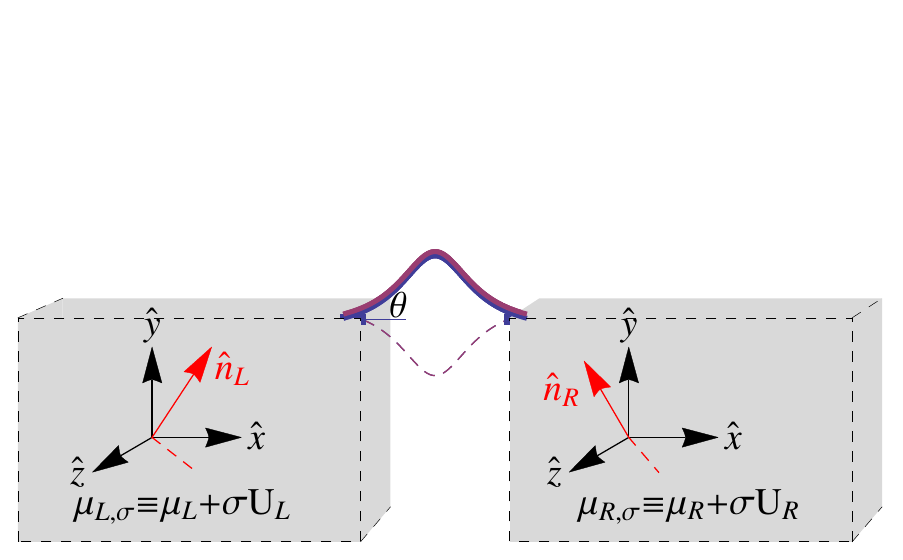}
\caption{A curved nanowire lying in the  $XY$ plane is coupled to two magnetically-polarized electronic reservoirs with arbitrarily-oriented	magnetization	axes,
$\hat{\bf n}_{L}$ and $\hat{\bf n}_{R}$. 
Externally-pumped spins give rise to  spin-dependent electrochemical potentials. The bending of the nanowire is specified by the angle it makes with $\hat{\bf x}$, with an instantaneous value $\theta$ around the equilibrium angle $\theta_{0}$. 
Reprinted figure with permission from R.  I.   Shekhter {\em et al.}, Phys. Rev. B {\bf 90}, 045401 (2014) \textcopyright 2014 by the American Physical Society.
}\label{Fig5}
\end{figure}

The setup we propose is illustrated  in Fig.~\ref{Fig5}. 
It comprises 
a nanowire 
bridging two leads, firmly coupled to the left and right electronic reservoirs,    held at spin-dependent electrochemical potentials,  
\begin{align}
\mu^{}_{L,\sigma}&=\mu^{}_{L}+\sigma U^{}_{L}\ ,\nonumber\\
\mu^{}_{R,\sigma}&=\mu^{}_{L}+\sigma U^{}_{R}\ ,
\label{MLR}
\end{align}
[generalizing Eq.~(\ref{mlr})].
The two bulk metals forming the reservoirs are each polarized along its own polarization axis, denoted by the unit vectors  $\hat{\bf n}_{L}$ and $\hat{\bf n}_{R}$, respectively. The wire vibrates in the $XY$ plane, such that the angle $\theta$ it makes with the $\hat{\bf x}-$axis oscillates around an equilibrium value,  $\theta_{0}$. An additional (weak) magnetic field,  applied along $\hat{\bf z}$, gives rise to an instantaneous Aharonov-Bohm effect.\cite{Glazman}

Since the electrodes are  magnetically-polarized, the density of states in each of them depends on both the internal exchange interaction and the external spin pumping as expressed by the energy split of the electrochemical potentials $U_{L,R}$ that determine the kinetic energy of the electrons participating in the transport. However, assuming the spin biases $U_{L(R)}$
to be much smaller than the Curie temperature in the magnetic leads, the
latter dependence is 
weak,  
and to lowest order in $U_{L,R}/\mu$, where $\mu=(\mu_{L}+\mu_{R})/2$ is the common chemical potential of the entire device, it may be neglected.

As shown in Appendix \ref{vib} [see in particular Eq.~(\ref{SSD})],  
 the spin-resolved particle currents emerging from the left and the right electrodes are [generalizing Eq.~(\ref{SR}) to include spin-dependent densities of states in the two bulky reservoirs]
\begin{align}
&-I^{}_{L,\sigma}=2\pi{\cal N}^{}_{L,\sigma}\sum_{\sigma '}{\cal N}^{}_{R,\sigma '}\sum_{n,n'=0}^{\infty}
P(n){\cal T}^{}_{nn',\sigma\sigma '}\nonumber\\
&\times
(1-e^{\beta(\mu^{}_{L,\sigma}-\mu^{}_{R,\sigma '})})\frac{\mu^{}_{L,\sigma}-\mu^{}_{R,\sigma '}+(n'-n)\omega^{}
_{0}
}{e^{\beta[\mu_{L,\sigma}-\mu_{R,\sigma '}+(n'-n)\omega^{}_{0}]}-1}\ ,\label{ils}
\end{align}
and 
\begin{align}
&-I^{}_{R,\sigma '}=2\pi{\cal N}^{}_{R,\sigma '}\sum_{\sigma }{\cal N}^{}_{L,\sigma }\sum_{n,n'=0}^{\infty}
P(n){\cal T}^{}_{nn',\sigma\sigma '}\nonumber\\
&\times
(e^{\beta(\mu^{}_{L,\sigma}-\mu^{}_{R,\sigma '})}-1)\frac{\mu^{}_{L,\sigma}-\mu^{}_{R,\sigma '}+(n'-n)\omega^{}_{0}
}{e^{\beta[\mu_{L,\sigma}-\mu_{R,\sigma '}+(n'-n)\omega^{}_{0}]}-1}\ .\label{irs}
\end{align}
Particle number is conserved, as can be seen by adding together Eq.~(\ref{ils}) summed over $\sigma$ and Eq.~(\ref{irs}) summed over $\sigma '$.

The spin indices of the matrix element squared    forming the transmission,  ${\cal T}$, in Eqs. (\ref{ils}) and (\ref{irs}) deserve some caution:  the quantization axes of the magnetization in the two electronic reservoirs are generally different (see Fig.~\ref{Fig5}), and they both may differ from the quantization axis which is used to describe the Rashba interaction on the nanowire. Specifying the quantization axis in the left  (right) reservoir  by the angles $\theta_{L}$ ($\theta_{R}$) and $\varphi_{L}$ ($\varphi_{R}$), then
\begin{align}
{\cal T}^{}_{nn',\sigma\sigma '}=|W^{}_{0}|^{2}|\langle n|[{\cal S}^{\dagger}_{R}e^{-i\psi^{}_{R}}e^{-i\psi^{}_{L}}{\cal S}^{}_{L}]^{}_{\sigma '\sigma}|n'\rangle|^{2}
\ , \label{T}
\end{align}
where the rotation transformations ${\cal S}_{L(R)}$ are given by
\begin{align}
&{\cal S}^{}_{L(R)}=\nonumber\\
&\left [\begin{array}{cc}
e^{-i\frac{\varphi^{}_{L(R)}}{2}}\cos \frac{\theta^{}_{L(R)}}{2}
&
e^{-i\frac{\varphi^{}_{L(R)}}{2}}\sin \frac{\theta^{}_{L(R)}}{2}
\\
e^{i\frac{\varphi^{}_{L(R)}}{2}}\sin
 \frac{\theta^{}_{L(R)}}{2}
&
-e^{i\frac{\varphi^{}_{L(R)}}{2}}\cos \frac{\theta^{}_{L(R)}}{2}
\end{array}\right ]\ .\label{slr}
\end{align}
For instance, when the quantization axes in both electrodes are identical, $\hat{\bf n}_{L}=\hat{\bf n}_{R}$, 
then ${\cal S}_{L}={\cal S}_{R}$ just rotates the direction of the  quantization axis of the 
Rashba interaction.

{\em The linear-response regime. }
In the linear-response regime,  the spin-resolved particle currents, Eqs.  (\ref{ils}) and (\ref{irs})
become
\begin{align}
&I^{}_{L,\sigma}=2\pi{\cal N}^{}_{L,\sigma }\sum_{\sigma '}{\cal N}^{}_{R,\sigma '}(\mu^{}_{L,\sigma }-\mu^{}_{R,\sigma '}){\cal A}^{}_{\sigma\sigma'}\ ,\nonumber\\
&I^{}_{R,\sigma '}=2\pi{\cal N}^{}_{R,\sigma '}\sum_{\sigma }{\cal N}^{}_{L,\sigma }(\mu^{}_{R,\sigma ' }-\mu^{}_{L,\sigma }){\cal A}^{}_{\sigma\sigma'}\ ,
\label{lr}
\end{align}
with the transmission
\begin{align}
{\cal A}^{}_{\sigma\sigma '}&=\sum_{n=0}^{\infty}
P(n){\cal T}^{}_{nn,\sigma\sigma '} \nonumber\\
&+\sum_{\stackrel{nn'=0}{n\neq n'}}^{\infty}P(n){\cal T}^{}_{nn',\sigma\sigma '}
\frac{
(n'-n)\beta\omega^{}_{0}
}{e^{(n'-n)\beta\omega^{}_{0}}-1}\ .
\label{a}
\end{align}
The first term in Eq.~(\ref{a}) gives the contribution to the spin-resolved transport  from the elastic tunneling processes.
The second is due to the inelastic processes, and is active at finite temperatures.

Our final expressions for the charge currents are then
\begin{align}
eI^{}_{L}&\equiv e\sum_{\sigma}I^{}_{L,\sigma}=e(\mu^{}_{L}-\mu^{}_{R}){\cal C}_{1}-eU^{}_{R}{\cal C}^{}_{3}+eU^{}_{L}{\cal C}^{}_{2}\ ,
\label{II}
\end{align}
with $eI_{R}\equiv e\sum_{\sigma '}I_{R,\sigma '}=-eI_{L}$. The spin currents emerging from the left and right reservoirs are
\begin{align}
I_{L}^{\rm spin}\equiv \sum_{\sigma}\sigma I^{}_{L,\sigma}=
(\mu^{}_{L}-\mu^{}_{R}){\cal C}^{}_{2}-U^{}_{R}{\cal C}^{}_{4}+U^{}_{L}{\cal C}^{}_{1}\ ,\nonumber\\
I_{R}^{\rm spin}=\sum_{\sigma'}\sigma' I^{}_{R,\sigma' }=
(\mu^{}_{R}-\mu^{}_{L}){\cal C}^{}_{3}+U^{}_{R}{\cal C}^{}_{1}-U^{}_{L}{\cal C}^{}_{4}\ .\label{IIS}
\end{align}
In Eqs. (\ref{II}) and (\ref{IIS}) we have introduced the  linear-response transport coefficients
\begin{align}
{\cal C}^{}_{1}&=2\pi\sum_{\sigma\sigma '}{\cal N}^{}_{L,\sigma}{\cal A}^{}_{\sigma\sigma '}{\cal N}^{}_{R,\sigma '}\ ,\nonumber\\
{\cal C}^{}_{2}&=2\pi\sum_{\sigma\sigma '}{\cal N}^{}_{L,\sigma}\sigma{\cal A}^{}_{\sigma\sigma '}{\cal N}^{}_{R,\sigma '}\ ,
\nonumber\\
{\cal C}^{}_{3}&=2\pi\sum_{\sigma\sigma '}{\cal N}^{}_{L,\sigma}{\cal A}^{}_{\sigma\sigma '}\sigma '{\cal N}^{}_{R,\sigma '}\ ,\nonumber\\
{\cal C}^{}_{4}&=2\pi\sum_{\sigma\sigma '}{\cal N}^{}_{L,\sigma}\sigma{\cal A}^{}_{\sigma\sigma '}\sigma '{\cal N}^{}_{R,\sigma '}\ ,\label{C}
\end{align}
giving  the various transmission probabilities of the junction. \cite{com2}

{\em The Onsager relations.}
As mentioned, there is a certain analogy between the configuration studied here and that of thermoelectric transport. In order to further pursue this point we consider the entropy production in our device, assuming that the spin imbalance in each of the two reservoirs does not vary with time and that all parts of the setup are held at the same temperature $T$. Under these circumstances  the entropy production, $\dot{S}$,  is
\begin{align}
T\dot{S}&=\sum_{\sigma}\mu^{}_{L,\sigma}I^{}_{L,\sigma}+\sum_{\sigma '}\mu^{}_{R,\sigma '}I^{}_{R,\sigma '}
\nonumber\\
&=I^{}_{L}(\mu^{}_{L}-\mu^{}_{R})+U^{}_{L}I^{\rm spin}_{L}+U^{}_{R}I^{\rm spin}_{R}\ ,\label{sd}
\end{align}
where the various currents are given in Eqs. (\ref{II}) and (\ref{IIS}).
Obviously, the first term on the right-hand side of Eq.~(\ref{sd}) is the dissipation due to Joule heating. The other two terms describe the dissipation involved with  the spin currents.

The entropy production may be presented as a scalar product of the vector of driving forces (the ``affinities"), $\{ V\equiv (\mu_{L}-\mu_{R})/e, U_{L},U_{R}\}$  and the resulting currents, $\{eI^{}_{L},I^{\rm spin}_{L},I^{\rm spin}_{R}\}$. In the linear-response regime 
these two vectors are related to one another by a (3$\times$3) matrix ${\cal M}$, 
\begin{align}
\left [\begin{array}{c}eI^{}_{L} \\I^{\rm spin}_{L} \\I^{\rm spin}_{R}\end{array}\right ]
={\cal M}\left [\begin{array}{c}V \\U^{}_{L} \\U^{}_{R}\end{array}\right ]\label{22}
\end{align}
with
\begin{align}
{\cal M}=\left [\begin{array}{ccc}
e^{2}{\cal C}^{}_{1}&e{\cal C}^{}_{2}&-e{\cal C}^{}_{3} \\    \  e{\cal C}^{}_{2}&\ \ {\cal C}^{}_{1}&-{\cal C}^{}_{4}\\   
-e{\cal C}^{}_{3}&-{\cal C}^{}_{4}&\ \ {\cal C}^{}_{1}\end{array}\right ]\ .
\label{23}
\end{align}
The matrix ${\cal M}$ contains the transport coefficients which do not depend on the driving forces.
One notes that this matrix obeys the Onsager reciprocity relations: reversing the sign of the magnetic field, i.e., inverting the sign of the  Aharonov-Bohm phases $\phi^{}_{L}$ and $\phi_{R}$ [see Eqs. (\ref{plr}) and (\ref{T})],  and simultaneously  interchanging the vibration states indices  $n$ with $n'$ and the spin indices  $\sigma$ with $\sigma'$ in Eqs. (\ref{T}) and (\ref{a}) leaves all off diagonal terms in the  matrix ${\cal M}$ unchanged.

{\em The transport coefficients.} The full calculation of the transmission  matrix ${\cal A}$ that determines the transport coefficients ${\cal C}_{i}$ [see Eqs. (\ref{a}) and (\ref{C})] is quite complicated, and requires a numerical computation. 
When the coupling of the charge carriers to the vibrational modes of the wire is weak, one may obtain an approximate expression by exploiting the different magnitudes  that coupling takes in  the magnetic Aharonov-Bohm phase and  in the Rashba one. In order to see this, it is expedient to 
present the phase factors in the transmission amplitude [see Eq.~(\ref{plr})] in the form
\begin{align}
\exp(-i\psi_{R})\exp(-i\psi_{L})\equiv e^{-i\phi}(A+i{\bf V}\cdot\sig)\ ,\label{p1n}
\end{align}
where $A$ and ${\bf V}$ are functions of the instantaneous bending angle $\theta$,  Eq.~(\ref{tet}), 
\begin{align}
&A=1-2\cos^{2}(\theta)\sin^{2}(k^{}_{\rm so} d/2)\ ,\nonumber\\
&{\bf V}=\{ 0,\cos (\theta)\sin(k^{}_{\rm so} d),-\sin (2\theta)\sin^{2}(k^{}_{\rm so} d/2)\}\ , \nonumber\\
&A^{2}_{}+{\bf V}\cdot{\bf V}=1\ , \label{p2n}
\end{align}
and $\phi=\phi_{L}+\phi_{R}$ is the instantaneous Aharonov-Bohm flux in units of the flux quantum $\Phi_{0}$.
The components of the spin-orbit vector ${\bf V}$ are given in the coordinate axes depicted in Fig.~\ref{Fig5}.

The effect of the electron-vibration interaction on the Rashba coupling is of  the order of the zero-point amplitude of the vibrations divided by the wire length, $a_{0}/d$. 
On the other hand, using Eq.~(\ref{tet}),  one finds that the Aharonov-Bohm phase, $\phi=-
[\pi Bd^{2}/(4\Phi_{0})]\sin(2\theta)$ ($B$ is the strength of the magnetic field), 
 is
\begin{align}
\phi \approx -\frac{\pi Bd^{2}}{4\Phi^{}_{0}}\sin (2\theta^{}_{0})-\frac{\pi a^{}_{0}d B}{2\Phi^{}_{0}}\cos (\theta^{}_{0})\cos (2\theta^{}_{0})(b+b^{\dagger})\ .
\end{align}
(The creation and destruction operators of the vibrations are denoted $b^{\dagger}$ and $b$.)
The dynamics of the Aharonov-Bohm flux is thus determined by the flux enclosed in an area of order $a_{0}d$  divided by the flux quantum. The latter ratio can be significantly larger than $a_{0}/d$. 
For instance, the length of a single-walled carbon nanotube is about $d=1\,\mu$m, while the vibrations' zero-point amplitude is estimated to be $10^{-5}\,\mu$m. This  leads to $a_{0}/d\approx 10^{-5}$, while $(Ba_{0}d)/\Phi_{0}$ is of the order of $10^{-2}$ for magnetic fields of the order of a few Teslas (at which the effect of the magnetic field on the transport through the Rashba weak link becomes visible).

The disparity between  the way the electron-vibration coupling affects the Rashba phase factor and the manner by which it dominates the magnetic one  results in a convenient (approximate) form for the transmission matrix ${\cal A}$, \cite{Shekhter.2013.2014,com2}    
\begin{align}
{\cal A}=G(T,B)\left [\begin{array}{cc}{\cal A}^{}_{\rm d}&{\cal A}^{}_{\rm nd}\\ {\cal A}^{}_{\rm nd}&{\cal A}^{}_{\rm d}\end{array}\right ]\ .\label{AM}
\end{align}
The  conductance  $G(T,B)$  (divided by $e^{2}$) derived in Ref.~\onlinecite{Glazman} [see  Eqs.~(\ref{GSP}) and (\ref{G(B)}) for the definition, and  Eq.~(\ref{LIMG}) for the limiting behaviors], gives  the transmission of the junction in the absence of the Rashba interaction; it depends on the temperature and on the perpendicular magnetic field.

The spin-dependent part of the transmission is given by the matrix in Eq.~(\ref{AM}), 
\begin{align}
{\cal A}^{}_{\rm d}+{\cal A}^{}_{\rm nd}&=1\ ,\nonumber\\
{\cal A}^{}_{\rm d}-{\cal A}^{}_{\rm nd}&=(A_{0}^{2}-V_{0}^{2})\hat{\bf n}^{}_{L}\cdot\hat{\bf n}^{}_{R}+2A^{}_{0}{\bf V}^{}_{0}\cdot\hat{\bf n}^{}_{L}\times\hat{\bf n}^{}_{R}\nonumber\\
&+2({\bf V}^{}_{0}\cdot\hat{\bf n}^{}_{L})({\bf V}^{}_{0}\cdot\hat{\bf n}^{}_{R})\ .\label{ADND}
\end{align}
Here $A_{0}$ and ${\bf V}_{0}$ are given by the values of  $A$ and ${\bf V}$ defined in Eqs. (\ref{p2n}) at equilibrium, i.e., when the angle $\theta$ there is replaced by $\theta_{0}$.
Their physical meaning  is explained below: ${\cal A}_{\rm nd}=\sin^{2}(\gamma)$, where $\gamma$ is the twisting angle of the charge carriers' spins, and ${\cal A}_{\rm d}=\cos^{2}(\gamma )$.

Using the explicit expression (\ref{AM}) for the transmission matrix ${\cal A}$
it is straightforward to find the transport coefficients ${\cal C}_{i}$. 
Retaining only terms linear in the difference between the densities of states of the two spin orientations, we obtain
\begin{align}
{\cal C}^{}_{1}+{\cal C}_{4}&\approx 8\pi G(T,B) {\cal A}^{}_{\rm d}{\cal N}^{}_{L}{\cal N}^{}_{R}\approx {\cal C}_{2}+{\cal C}^{}_{3}\ ,\nonumber\\
{\cal C}^{}_{1}-{\cal C}_{4}&\approx 8\pi G(T,B){\cal A}^{}_{\rm nd}{\cal N}^{}_{L}{\cal N}^{}_{R}\ ,\nonumber\\
{\cal C}^{}_{2}-{\cal C}^{}_{3}&=4\pi G(T,B){\cal A}^{}_{\rm nd}({\cal N}^{}_{L,\up}{\cal N}^{}_{R,\down}-{\cal N}^{}_{L,\down}{\cal N}^{}_{R,\up})\ ,\label{C1}
\end{align}
where ${\cal N}_{L,R}$ is the total density of states of each electronic reservoir (summed over the two spin directions).
Glancing at Eq.~(\ref{II})
for the charge current,  and taking into account the first of Eqs. (\ref{ADND}), shows that the conductance, $G$,  of the junction is independent of the spin-orbit interaction, and is given by 
\begin{align}
G=4\pi e^{2}_{}{\cal N}_{L}{\cal N}^{}_{R}G(T,B)\ .
\label{GC}
\end{align}

{\em Rashba twisting.}
When the junction is not subjected to a perpendicular magnetic field and the charge carriers passing through it do not collect an Aharonov-Bohm phase due to it, one may safely ignore the effect of the quantum flexural nano-vibrations of the suspended wire. \cite{Shekhter.2013.2014}
The scattering of the 
electrons' momentum, caused by the spatial constraint of their orbital motion inside the nanowire, also induces 
scattering of the electronic spins. The latter results from the SO Rashba interaction located at the wire. Consequently,  an electronic wave having  a definite  spin projection  on the magnetization vector of  the lead from which it emerges, is not a spin eigenstate in the other lead.   
Thus, a pure spin state $|\sigma\rangle$ in one lead becomes a mixed spin state 
in the other, 
\begin{align}
|\sigma\rangle\Rightarrow
\alpha_{1}|\sigma\rangle+\alpha_{2}|\overline{\sigma}\rangle\ ,
\label{amps}
\end{align}
with probability amplitude $\alpha_{1}$ to remain in  the original state, and probability amplitude  $\alpha_{2}$ for a spin flip ($\overline{\sigma}=-$$\sigma$).
During the propagation   through the weak link the spins
of the charge carriers are twisted, 
as is described by the transmission amplitude [see Eq.~(\ref{p1n})],
$A^{}_{0}+i{\bf V}^{}_{0}\cdot\sig$.  
It follows that the probability amplitude for a spin flip, $\alpha_{2}$, 
is  given by
\begin{align}
\alpha^{}_{2}=[{\cal S}^{\dagger}_{R}(A^{}_{0}+i{\bf V}^{}_{0}\cdot\sig){\cal S}^{}_{L}]^{}_{\sigma\overline{\sigma}}\ ,
\end{align}
with  ${\cal S}_{L,R}
$ given in Eq.~(\ref{slr}). The Rashba  twisting angle, $\gamma$,  can now be defined by
\begin{align}
\alpha^{}_{2}=\sin(\gamma )e^{i\delta}\ ,
\end{align}
with
\begin{align}
|\alpha^{}_{2}|^{2}=\sin^{2}(\gamma)={\cal A}_{\rm nd}\ , \label{AG}
\end{align}
yielding a clear physical meaning to the transmissions ${\cal A}_{\rm d}$ and ${\cal A}_{\rm nd}$ [see Eqs. (\ref{ADND})].
The physical quantities depend only on the relative phase between $\alpha_1$ and $\alpha_2$. Therefore, we choose  $\alpha_1=\cos\gamma$.
It is then easy to check that the average of the vector $\sig$ in the state of Eq.~(\ref{amps}) is equal to $\{\sin(2\gamma)\cos(\delta),\sin(2\gamma)\sin(\delta),\cos(2\gamma)\}$. This vector is rotated by the angle $2\gamma$ relative to its direction in the absence of the spin-orbit interaction. This rotation of the electronic moments in each of the two leads is  a ``twist" of the spins.  It is distinct from simple spin precession since the axis of this precession changes its direction during the electronic motion along the curved trajectory.

In the simplest configuration of parallel magnetizations in both electrodes, i.e., 
\begin{align}
\hat{\bf n}_{L}^{}=\hat{\bf n}^{}_{R}\equiv \hat{\bf n}\ ,
\end{align}
Eqs. (\ref{ADND}) yield
\begin{align}
\sin (\gamma )=[V_{0}^{2}-(\hat{\bf n}\cdot{\bf V}^{}_{0})^{2}]^{1/2}
\ .\label{gama}
\end{align}
Interestingly enough, in this simple configuration $\sin (\gamma)$ is determined by the component of the Rashba   vector ${\bf V}_{0}$ normal to the quantization axis of the magnetization in the electrodes.
Mechanically manipulating the bending angle that determines the direction of the Rashba vector ${\bf V}_{0}$,  one may control the twisting angle $\gamma$. Note also that had the vectors $\hat{\bf n}_{L}$ and $\hat{\bf n}_{R}$ been antiparallel to one another
then $\sin (\gamma )=[1-V_{0}^{2}+(\hat{\bf n}\cdot{\bf V}^{}_{0})^{2}]^{1/2}$.

An even more convenient way to monitor 
the twisting effect
may be realized by studying  the spintro-voltaic 
effect in an open circuit, i.e., when the total charge current vanishes. One then finds that the spin-imbalanced populations in the electrodes give rise to an electric voltage, $V_{sv}$.  Assuming  that the spin imbalances in the two reservoirs are identical, i.e., $U_{L}=U_{R}\equiv U$, Eq.~(\ref{II}) yields
\begin{align}
V^{}_{sv}=\frac{{\cal C}^{}_{3}-{\cal C}^{}_{2}}{{\cal C}^{}_{1}}U\ .\label{vsv}
\end{align}
The ratio of the voltage created by the spin imbalance, $V_{sv}$, to the amount of spin imbalance in the electrodes (expressed by $U$) can be found upon using Eqs. (\ref{C1}), in conjunction with Eqs. (\ref{ADND}) and  (\ref{AG}), 
\begin{align}
V^{}_{\rm sv}=\sin^{2}(\gamma) \frac{{\cal N}^{}_{L\up}{\cal N}^{}_{R\down}-{\cal N}^{}_{L\down}{\cal N}^{}_{R\up}}{{\cal N}^{}_{L}{\cal N}^{}_{R}}U\ .
\end{align}
The voltage generated by the Rashba interaction gives directly the twisting angle; the proportionality between $V_{\rm sv}/U$ and $\sin^{2}(\gamma) $ is the magnetic mismatch parameter of the junction.

The twisting angle $\gamma$ 
determines also the  various spin conductances of the junction. From Eqs. (\ref{22}) and (\ref{23}) we find \cite{com6}
\begin{align}
&I^{\rm spin}_{L}+I^{\rm spin}_{R}=eV({\cal C}_{2}-{\cal C}^{}_{3})+(U^{}_{L}+U^{}_{R})({\cal C}_{1}^{}-{\cal C}^{}_{4})\ .
\nonumber\\
&=2\frac{G}{e^{2}}
{\cal A}^{}_{\rm nd}\Big (U^{}_{L}+U^{}_{R}+eV
\frac{{\cal N}^{}_{L,\up}{\cal N}^{}_{R,\down}-{\cal N}^{}_{L,\up}{\cal N}^{}_{R,\down}}{2{\cal N}^{}_{L}{\cal N}^{}_{R}}\Big )\ ,
\label{41}
\end{align}
where $G$ is the charge conductance, Eq.~(\ref{GC}), and we have made use of Eqs. (\ref{C1}) for the ${\cal C}$'s.
One now observes that both the spin conductance, $G^{\rm spin}$ (normalized by the charge conductance)
\begin{align}
G^{\rm spin}_{}=\frac{I^{\rm spin}_{L}+I^{\rm spin}_{R}}{(U^{}_{L}+U^{}_{R})G/e^{2}}\Big |^{}_{V=0}\ ,
\label{NSC}
\end{align}
and the cross spin conductance, $G^{\rm spin}_{\times}$, (again normalized by the charge conductance),
\begin{align}
G^{\rm spin}_{\times}=\frac{I^{\rm spin}_{L}+I^{\rm spin}_{R}}{eV
G/e^{2}}\Big |^{}_{U_{L}^{}=U_{R}^{}=0}\ ,
\label{43}
\end{align}
are determined by ${\cal A}_{\rm nd}$, that is by the twisting angle $\gamma$, Eq.~(\ref{AG})
(the second requires the asymmetry  in the spin-resolved densities of states).

For parallel magnetizations in the leads, the twisting angle [see Eq.~(\ref{gama})]
depends solely on the SO coupling and on the equilibrium value of the bending angle. The spin twisting disappears  for any direction of the polarizations in the leads at $\theta_{0}=\pi/2$. This can be easily understood within  a classical picture for the spin rotation caused by the Rashba interaction. 
The spin evolution of the tunneling electron can be regarded as a rotation around an axis given by the vectorial product of the velocity and the electric field (directed along $\hat{\bf z}$ in our configuration). At
this value of $\theta_{0}$ the tunneling trajectory is oriented along  $\hat{\bf y}$ (because then $x_{R}=x_{L}=0$)       
and so the electron ``rattles" back and forth along $\hat{\bf y}$. This leads to a cancellation of the Rashba contribution to the tunneling phase [see Eq.~(\ref{plr})]. 
The other special case is when the wire is not bended, i.e., $\theta_{0}=0$. The spin twisting for leads' magnetizations along $\hat{\bf y}$ vanishes,  while for devices with ferromagnetic magnetizations along the  $\hat{\bf x}-$ or $\hat{\bf z}-$directions it reaches its maximal value, $\sin(k_{\rm so} d)$. The reason for this has also to do with the orientation of the spin rotation-axis. At small values of $\theta_{0}$ the electronic 
trajectory is primarily along $\hat{\bf x}$. Then, when the spin of the incident electron is directed along $\hat{\bf y}$  it is parallel to the rotation axis and no rotation is taking place. In contrast, when the spin of the incident electron is oriented along $\hat{\bf x}$ or $\hat{\bf z}$, it is perpendicular to the rotation axis, leading to a full rotation.

Thus,  one can have spintro-electric functionalities if one uses a vibrating  suspended weak link, with both a magnetic flux and an (electric field dependent) Rashba spin-orbit interaction. The twisting of the electronic spins as they move between the (spin-polarized) electrodes can be manipulated by the bias voltage, the bending of the weak link wire,  and the polarizations  in the electrodes.
The twisting angle, which determines the probability amplitude of the Rashba splitting,  can be measured electrically through a spintro-voltaic effect.

\section{Spin polarization of Cooper pairs in spin-orbit-active superconducting weak links}
\label{Josephson}

A remarkable consequence of superconductivity, known as  the proximity effect, allows a supercurrent to flow between two superconductors connected by a non-superconducting material of  a finite width. This phenomenon was actively studied during several decades starting with the pioneering prediction by Josephson \cite{Josephson} in the early 1960's that  a non-dissipative current may flow through a tunnel junction formed by a layered superconductor-insulator-superconductor (S-I-S) structure. A number of other so-called superconducting weak links, involving a normal metal, a quantum dot and various micro-constrictions, have been studied theoretically and experimentally during the past decades in order to explore and exploit this  phenomenon. 

A phase transition to a superconducting state is accompanied by the formation of a new ground state, which is different from the standard metal ground state and which can be viewed as a condensate of paired electrons (Cooper pairs). The pairing is supported by an indirect attractive inter-electronic interaction usually described in terms of a ``pairing potential". In inhomogeneous superconductors this potential is coordinate dependent and can, in particular, be suppressed in a layer perpendicular to the direction of the superconducting current. In this case electron pairs that carry the supercurrent are  injected into a region where pairing is no longer supported by a pairing potential. However, the coherent properties of the electrons established inside the superconducting injector can be preserved for a certain distance (the superconducting coherence length),
which allows a non-dissipative current to flow through a non-superconducting layer of a sufficiently small width. Nevertheless, the spatial segregation of the Cooper pairs from the pairing potential responsible for their stability suggests a unique way for manipulating the Cooper pairs during their propagation through a superconducting weak link.

Consider,  for example,  the well-known fact that electrons, which form a Cooper pair in a conventional (singlet BCS) superconductor, are in time-reversed quantum states and therefore their spins are aligned in opposite directions so that the pair as a whole carries no spin. This spin ordering can be distorted inside the weak link, which allows for an intentional ``spin design" to be achieved by means of a superconducting weak link. 
In this Section we describe a particular mechanism for this kind of spin design, {\em viz.} an SO interaction localized to the non-superconducting weak link as presented in Ref. \onlinecite{Shekhter.2016b}

We show that the  splitting of the spin state of the paired electrons that carry the Josephson current
may transform the spin-singlet Cooper pairs into a coherent mixture of singlet and triplet spin states.
This mixture gives rise to  interference  between the channel in which both electrons preserve their spins and the channel where they are
flipped. The resulting  interference  pattern, that appears in  the Josephson current but does not show up in the normal-state transmission of the junction, allows for electrical and mechanical control of the Josephson current between two spin-singlet superconductors; it corresponds to
a new type of ``spin-gating" \cite{Shekhter.2016} of superconducting ``weak links".

\begin{figure}[htp]
\vspace{0.cm} \centerline {\includegraphics[width=11cm]{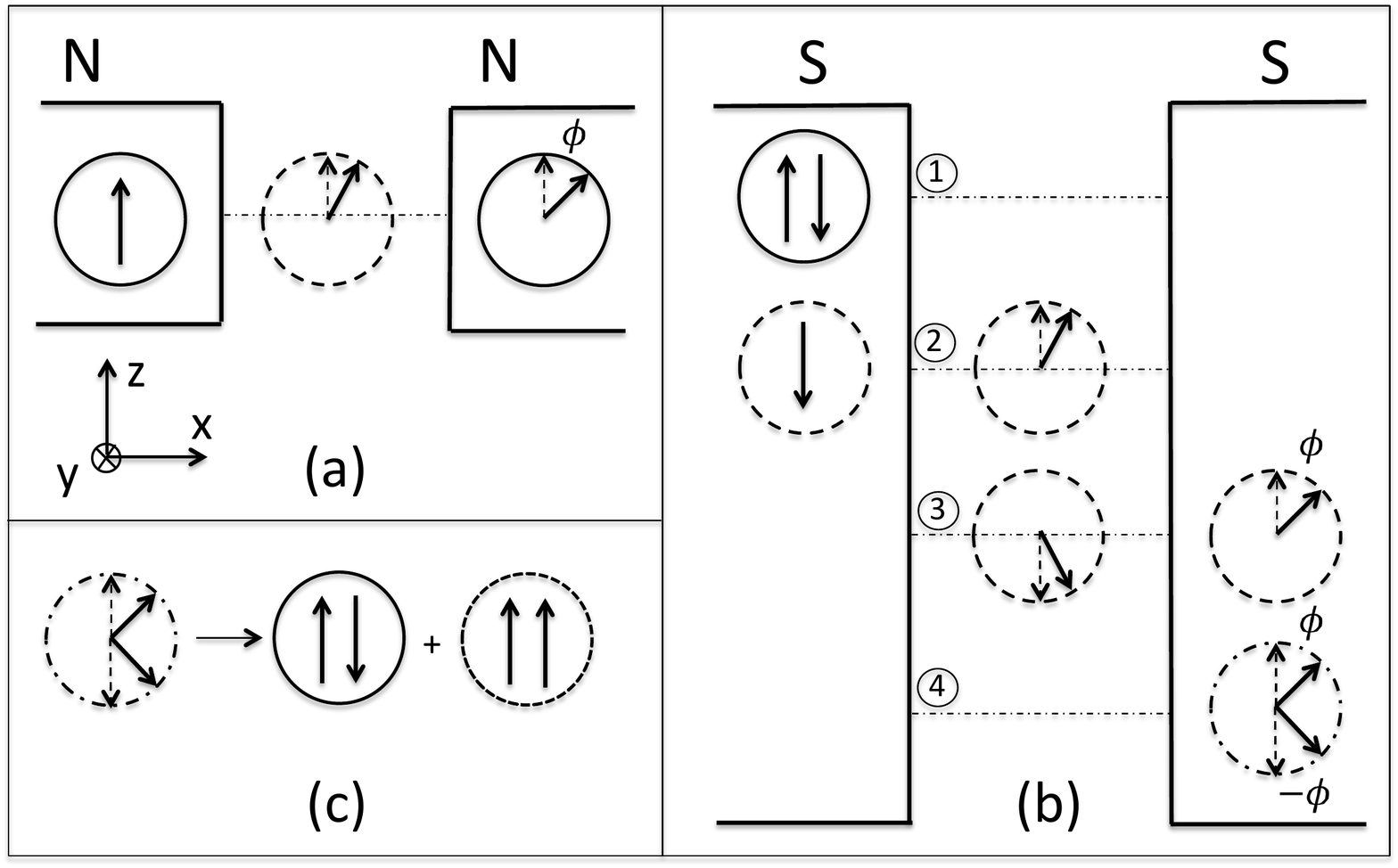}}
\caption{Schematic illustrations of the lowest-order perturbation expansion steps for tunneling (in the Coulomb blockade regime) through a straight nanowire weak link 
subjected to the Rashba spin-orbit (SO) interaction
caused by an electric field along  $\hat{\bf z}$. In a semiclassical picture, the spin of each electron (denoted by an arrow) is rotated in the $XZ-$plane 
as it goes through the link.
(a)
Single electron tunneling from one normal metal to another, via an intermediate (rotating) state (dashed circle).  When the electron enters the second normal metal, its spin has been rotated. 
(b) Sequential tunneling in four steps of a Cooper pair between two superconductors connected by the same weak link.
Because the two electrons that form the Cooper pair are in time-reversed states,
the SO interaction rotates their spins in opposite directions. (c) As they enter the second superconductor, the Cooper pairs are in a coherent mixture (dash-dotted circle) of a spin-singlet and a spin-triplet state. 
Inside this superconductor, this state is then
projected onto the singlet state (full circle). 
Reprinted figure with permission from R.  I.   Shekhter {\em et al.}, Phys. Rev. Lett. {\bf 116}, 217001 (2016) \textcopyright 2016 by the American Physical Society.}
\label{Fig6}
\end{figure}


To illustrate our calculation, Fig.~\ref{Fig6}
uses a semiclassical analogue of the quantum
evolution of the spin states of electrons which move between two bulk leads via a weak link, where they are subjected to the Rashba SO interaction. For simplicity we assume for now that the weak link is a straight 1D wire along the $\hat{\bf x}-$axis. The  SO interaction in the wire is due to an  electric field, which for the moment is assumed to point along $\hat{\bf z}$ and therefore corresponds to an effective SO-interaction-induced magnetic field  directed along  $\hat{\bf y}$.
Figure~\ref{Fig6}(a) illustrates  a single-electron  transfer from one normal metal to another.  Without loss of generality, we choose the $\hat{\bf z}-$axis  in spin space to be along the direction of the polarization of the electron in the first (left) metal.
Semiclassically, the spin of the injected electron rotates in the $XZ-$plane
as it passes through the wire.
As a result, the spins of the electrons that enter the second metal from the wire are rotated around the $\hat{\bf y}-$axis by an angle
proportional to the strength of the SO interaction and the length of the wire. This rotation depends on the direction of the ``initial" electron's polarization. It occurs only if the polarization has a component in the $XZ-$plane.
Quantum mechanically, the electron's spinor in the left metal is an eigenfunction of the Pauli spin-matrix $\sigma_z$, and the spinor of the outgoing electron is in a coherent superposition of spin-up and spin-down eigenstates of $\sigma_z$.

How can this picture be generalized to describe the transfer of the two electrons of a Cooper pair between two bulk superconductors?
The simplest case to consider, which we focus upon below, is when the  single-electron tunneling is Coulomb-blockaded throughout the wire. While the blockade can be lifted for one electron, double electron occupancy of the wire is suppressed, i.e.,
a Cooper pair is mainly transferred sequentially,
as shown in Fig.~\ref{Fig6}(b).
Each electron transfer is now accompanied by the  spin rotation shown in Fig.~\ref{Fig6}(a). However, since the two transferred electrons
are in time-reversed quantum states, the time evolution of their spins are reversed with respect to one another, and their rotation angles have opposite signs [step 4 in Fig.~\ref{Fig6}(b)].
This final state [Fig.~\ref{Fig6}(c)] can be expressed as a coherent mixture of a spin-singlet and a spin-triplet state, but only the former can enter into the second superconductor. As we show below, this projection onto the singlet causes a reduction of the Josephson current.

We consider a model where a Cooper pair is transferred
between superconducting source and drain leads via virtual states localized in a weak-link wire   [see Fig.~\ref{Fig7}(a)].
 The corresponding tunneling process, which
 supports multiple tunneling channels, was analyzed in detail in Ref.~\onlinecite{Gisselfalt.1996}. For simplicity, it is assumed throughout this section that the angle $\theta$ remains fixed, that is, the wire does not vibrate.  A significant simplification occurs in the Coulomb-blockade regime, defined by the inequality{
$E_{e}=E_{C}(N+1)-E_{C}(N)\gg |\Delta |$,
where $|\Delta|$  is the  energy gap parameter in the superconducting leads,  \cite{note} and $E_C(N)$  is the Coulomb energy of the wire when it contains $N$ electrons. In this regime,  tunneling channels requiring two electrons to be simultaneously localized in a virtual state in the wire can be neglected, and hence the tunneling processes are  sequential. Another simplification follows from our assumption that the length of the wire $d$ is short compared to the superconducting coherence length $\xi_0 \equiv \hbar v_{\rm F}/|\Delta|$,  \cite{note} so that the dependence of the matrix element for a single electron transfer on the electron energy in the virtual states can be ignored. A final simplification, facilitated by the device geometry,   concerns the conservation of the electrons' longitudinal momenta as they tunnel between the two leads.
In Fig.~\ref{Fig7}, the  wire ends are placed on top of the metal leads and are separated from them by thin but long tunneling barriers. Since the direction of tunneling is nearly perpendicular to the direction of the current along the wire, such a geometry is conducive to longitudinal momentum conservation. 
\cite{Eaves.2013}
\begin{figure}
\vspace{-0.5cm} \centerline {\includegraphics[width=10cm]{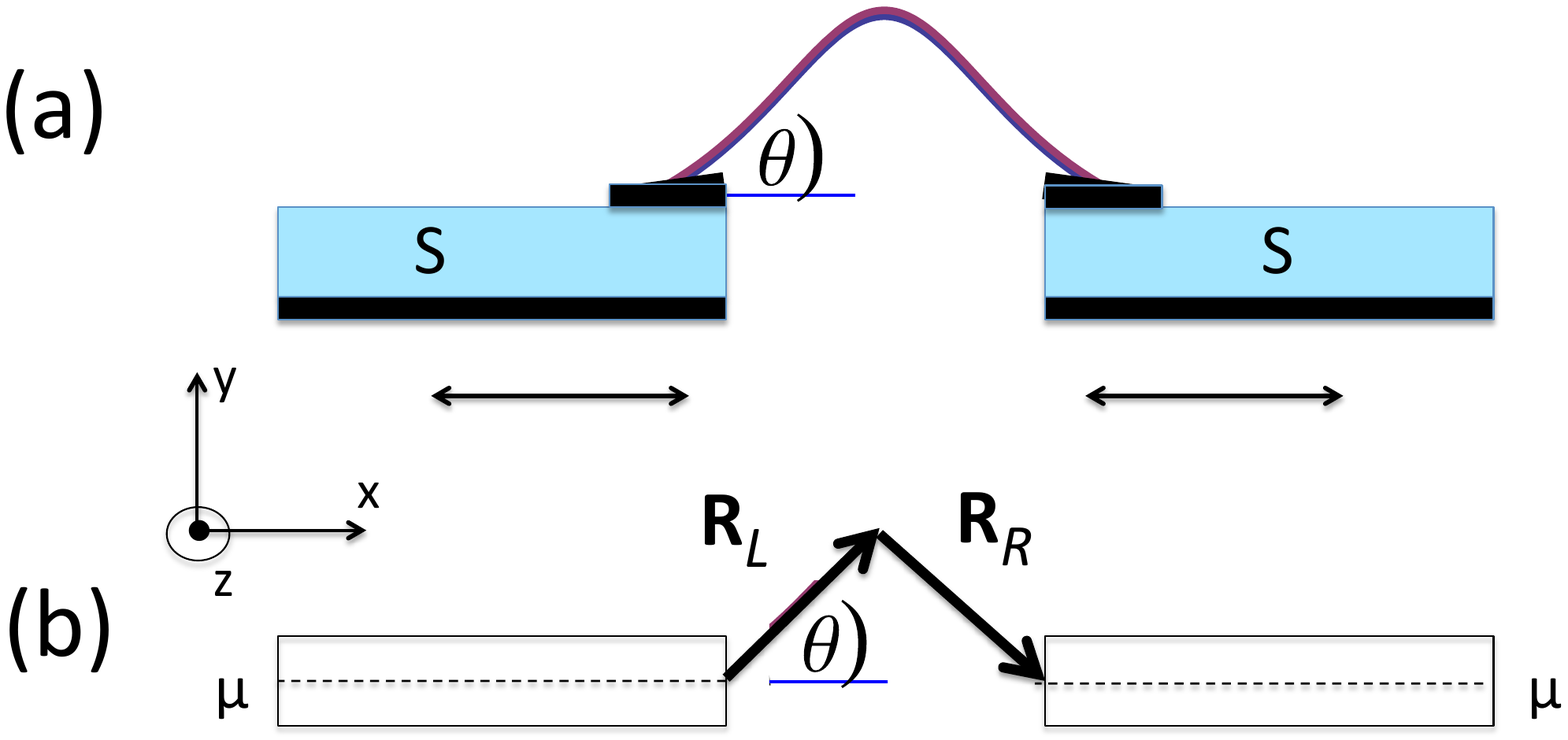}}
\vspace*{-2.2 cm}
\caption{(color online)
Sketch (a) and simplified model (b) of a
device that would allow the effects predicted in the text  to be studied. 
Reprinted figure with permission from R.  I.   Shekhter {\em et al.}, Phys. Rev. Lett. {\bf 116}, 217001 (2016) \textcopyright 2016 by the American Physical Society.}
\label{Fig7}
\end{figure}

These simple but realistic assumptions  allow us to describe the transfer of a Cooper pair between the two superconductors  in terms of
single-electron tunneling,
 as  given by the Hamiltonian (\ref{fullH}) [see also Eqs. (\ref{HT})-(\ref{delta})], with the tunneling matrix elements derived in Appendix \ref{app.tun}. 
We assume a weak link containing a bent wire [see Fig.~\ref{Fig7}(a)]. The actual calculations are done for the geometry shown in Fig.~\ref{Fig7}(b), where the weak link comprises two straight one-dimensional  wires, ${\bf R}_L$ and ${\bf R}_R$, of equal length $d/2$,  connected by a ``bend". The angles between these wires and the $\hat{\bf x}-$axis are $\theta$ and $-\theta$, respectively.

In the absence of the SO interaction, the supercurrent scales as the transmission of the junction when in the normal state.  \cite{Ambegaokar.Baratoff,Kulik}  As detailed in Appendix \ref{app.src},    the SO coupling modifies this transmission by  the factor  
Tr$\{{\cal W}{\cal W}^\dag\}$,  
where
\begin{align}
{\cal W} = e^{-i{k}_{\rm so}\sig\cdot {\bf R}^{}_{L}\times\hat{\bf n}}
e^{-i{k}_{\rm so}\sig\cdot {\bf R}^{}_{R}\times\hat{\bf n}},
\label{WLR}
\end{align}
and the trace is carried out in spin space.
When $k_{\rm so}$ vanishes, this  factor is simply 2, the spin degeneracy; i.e., the SO interaction does not affect the electric conductance (unless the junction allows for geometrically-interfering processes \cite{Ora.and.Amnon}).
The superconducting Josephson current is
\begin{align}
\frac{J(\varphi)}{J^{}_0(\varphi)} =\frac{1}{2}
\sum_\sigma\left[
\vert
{\cal W}_{\sigma\sigma}\vert^2 -\vert
{\cal W}_{\sigma\bar{\sigma}}\vert^2 \right]
=
\sum_\sigma\left[
\frac{1}{2}-\vert
{\cal W}_{\sigma\bar{\sigma}}\vert^2
 \right]\ ,
\label{eq:13}
\end{align}
where 
$J^{}_0(\varphi)\propto \sin(\varphi)$ is the equilibrium Josephson current in the absence
of the SO interaction,   \cite{Ambegaokar.Baratoff} and $\varphi$ is the superconducting phase
difference. 

Hence, the SO interaction modifies significantly the amplitude of the Josephson equilibrium current, while leaving the transmission of the junction in its normal state as
in the absence of this coupling. The matrix 
${\cal W}$, that determines these quantities,  
depends crucially on the direction $\hat{\bf n}$ of the electric field [see Eq.~(\ref{HAM})]. In  the configuration where
$\hat{\bf n}$ is normal to the plane of the junction, which is  described semiclassically in Fig.~\ref{Fig6}, 
$\hat{\bf n}\parallel\hat{\bf z}$, and then
\begin{align}
&{\cal W}
=
\big[\cos^{2}(k^{}_{\rm so}d/2)-
\sin^{2}(k^{}_{\rm so}d/2)
\cos (2\theta)\big]\nonumber\\
&+i\sig\cdot\big [\hat{\bf y}
\sin(k^{}_{\rm so}d)\cos(\theta)
+\hat{\bf z}\sin^{2}(k^{}_{\rm so}d/2)\sin (2\theta)
\big ]\ .
\label{WLR1}
\end{align}
In contrast, when the electric field is in the plane of the junction, e.g.,
$\hat{\bf n}=\hat{\bf y}$, we find
\begin{align}
{\cal W}=
\cos [k^{}_{\rm so}d\cos(\theta)]-i\sig\cdot\hat{\bf z}\sin [k^{}_{\rm so}d\cos(\theta)]\ .
\label{WLR2}
\end{align}
When the SO interaction is given by Eq.~(\ref{HAMstrain}), one finds that ${\cal W}$ of the strain-induced case has the same form as
Eq.~(\ref{WLR1}), except that $\hat{\bf y}$  is replaced by $\hat{\bf x}$. The resulting expressions for  the Josephson current and for the  normal-state transmission turn out to be the same as  for the
SO interaction of Eq.~(\ref{HAM}), with 
 $\hat{\bf n}\parallel\hat{\bf z}$.

 The matrix element
${\cal W}_{\sigma\bar{\sigma}}$
depends on the quantization axis of the spins. Choosing this axis to be along $\hat{\bf z}$,
 then when  $\hat{\bf n}\parallel\hat{\bf z}$    Eq.~(\ref{WLR1}) yields
({\it cf.} Fig.~\ref{Fig8})
 \begin{align}
\label{eq:17}
1 -2 \vert
{\cal W}_{\sigma\bar{\sigma}
}
\vert^2=1-2\cos^{2}(\theta)\sin^{2}(k^{}_{\rm so}d)\ .
\end{align}
In contrast,
 when the electric field is in the plane of the junction,
$\hat{\bf n}=\hat{\bf y}$, the matrix 
${\cal W}$
is diagonal [Eq.~(\ref{WLR2})],
$J(\varphi)=J_0(\varphi)$, and
the superconducting current is not affected by the spin dynamics.
Similar qualitative results are found  for all the directions of the spin quantization axis.
 For example, for spins polarized along $\hat{\bf y}$, one finds
$\vert 
{\cal W}_{\sigma\bar{\sigma}
}
\vert^2=\sin^{4}(k^{}_{\rm so} d/2)\sin^2(2\theta)$  when $\hat{\bf n}\parallel\hat{\bf z}$, while when    $\hat{\bf n}\parallel\hat{\bf y}$ it is
 $\vert
{\cal W}_{\sigma\bar{\sigma}
}\vert^2=\sin^2[k^{}_{\rm so}d \cos(\theta)]$. 
In most 
cases,
the splitting of the Cooper-pair spin state by the SO 
interaction reduces significantly 
the Josephson current through 
the superconducting weak link under consideration.

\begin{figure}
\vspace{0.cm} \centerline {\includegraphics[width=7.5cm]{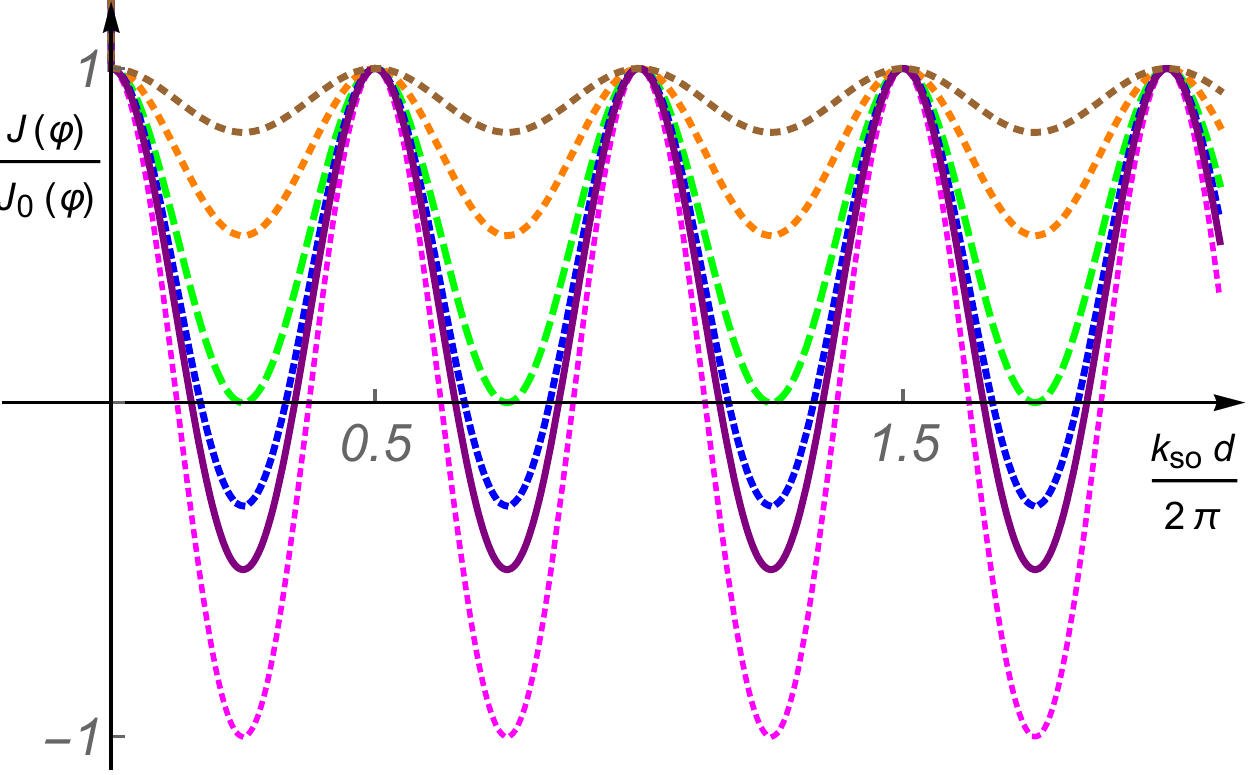}}
\caption{(color online)
The Josephson current $J(\varphi)$  divided by its value without the SO interaction, $J_{0}(\varphi)$,
for the SO interaction Eq. (\ref{HAM}), 
as a function
of $k_{\rm so} d/(2\pi)$. The largest amplitude is  for zero bending angle, $\theta=0$, decreasing   gradually for $\theta= \pi/6,\ \pi/5 ,\ \pi/4,\ \pi/3 ,$ and $\pi/2.5$.
Relevant values of $k_{\rm so}$ are estimated in Sec. \ref{conclusion}. 
Reprinted figure with permission from R.  I.   Shekhter {\em et al.}, Phys. Rev. Lett. {\bf 116}, 217001 (2016) \textcopyright 2016 by the American Physical Society.
}
\label{Fig8}
\end{figure}

Two features determine the magnitude of the effect for a given spin quantization axis in the leads (in addition to the strength $k_{\rm so}$ of the SO interaction and the length $d$ over which it acts). One is the extent to which the nanowire is bent 
($\theta$ in Fig.~\ref{Fig7}),  and the other is the orientation $\hat {\bf n}$ of the electric field responsible for the SO coupling  relative to the spin quantization axis. Both 
break spin conservation, which results in Rabi oscillations between the singlet  and triplet spin states of the (originally spin-singlet) Cooper pairs passing 
through the SO-active weak link.  The consequence is a spin splitting of the Cooper pairs that reach the second superconducting lead, where their spin state is projected onto the singlet state.
This splitting can result in a Josephson current that is an oscillatory function of the ``action" $k_{\rm so} d$ of the SO interaction  (Fig.~\ref{Fig8});  
the effect may be absent for special directions of the electric field.  Both results can 
be understood in terms of a semiclassical picture, 
Fig.~\ref{Fig6}.

As seen in Eq.~(\ref{eq:13}), the Josephson current can be written as a sum of two contributions. 
One, $\vert
{\cal W}_{\sigma\sigma}
\vert^2$, comes from a channel where the spin projections of the Cooper pair electrons, when leaving and entering 
the weak link, 
are identical; the other,
$\vert
{\cal W}_{\sigma\bar{\sigma}
}
\vert^2$, arises from  another channel, where the electron spins are flipped
during the passage. It is remarkable that the two contributions have opposite signs. This is due to  a Josephson tunneling ``$\pi$-shift"
caused by electronic spin flips 
(and is similar to the effect predicted for tunneling through a Kondo impurity \cite{Glazman.1989.Spivak.1991}). 
In particular,  a total cancellation of the Josephson current is possible when, e.g., 
$\theta = 0$ and $k_{\rm so} d = \pi/4$;  
in the limit $\theta=0$ and $k_{\rm so} d = \pi/2$ the Josephson current even changes its
sign.
This spin-orbit induced interference effect on the Josephson current is specific to a   weak link  subjected to SO interaction between superconductors. There is no such effect on the current through a single weak link connecting two normal metals.

According to Eq.~(\ref{eq:13}), none or both of the Cooper pair electrons must have flipped their spins as they leave the weak link in order to contribute to the Josephson current. This is because 
only spin-singlet Cooper pairs can enter the receiving $s$-type bulk superconductor. However,  single-flip processes, where only one of the two tunneling electrons flips its spin, are also possible results of injecting Cooper pairs into a Rashba weak link. Those processes correspond to a triplet component of the spin state of the transferred  pair,  and  can be viewed as
evidence for  {\em spin polarization of  injected Cooper pairs}. The triplet component  could be responsible for a spin-triplet proximity effect, 
\cite{Reeg.2015}
and would presumably contribute a spin supercurrent 
if higher-order tunneling processes were taken into account.

Thus, 
the supercurrent can be tuned by mechanical and electrical manipulations of the spin polarization of the Cooper pairs. 
In particular,  the Josephson current through an electrostatically-gated device becomes an oscillatory function of the gate voltage.
We emphasize that these results follow from the interference of two transmission channels, one where the spins of both members of a Cooper pair are preserved and one where they are both flipped, and that
this interference does not require any external magnetic field.
It is important, however, that those parts of the device where the superconducting pairing potential is non-zero and where the SO coupling is finite are spatially separated. To lowest order in the tunneling this separation prevents the superconductivity in the leads to have any effect on the dynamical spin evolution in the wire.

\section{Conclusions}
\label{conclusion}

In addition to the charge of electrons, their spin degree of freedom can also play an important role when nanometer-sized devices are used for electronics applications. The electron spin naturally comes into play if magnetic materials or external magnetic fields are used. However, even in non-magnetic materials the spin may couple strongly to an effective magnetic field induced by the spin-orbit (SO) interaction, which is a relativistic effect that couples the electron's spin degree of freedom to its orbital motion. Such a coupling, first discovered in bulk materials without spatial inversion symmetry, can be significantly enhanced in nanostructures where the screening of an electrostatic field is suppressed and   spatial inversion symmetry can be lifted by internal or external electric fields. The concept of an enhanced spin-orbit coupling in the vicinity of crystal surfaces, \cite{Vasko.1979, Rashba.1984} can therefore naturally be extended to nanostructures, where the surface to volume ratio can be rather high.

Carbon nanotubes and semiconductor wires 
seem particularly suitable to be used as  spin-splitters.
Measured Rashba  spin-orbit-coupling induced energy gaps in
InGaAs/InAlAs ($\Delta_{\rm so}=2\hbar v_{\rm F}k_{\rm so}\approx 5$~meV) \cite{Nitta.1997} and 
InAs/AlSb ($\Delta_{\rm so} \approx 4$~meV) \cite{Heida.1998} quantum wells
correspond to 
$k_{\rm so} \approx 4\times 10^6$ m$^{-1}$. 
The strain-induced SO energy gap for a carbon nanotube  is
$\Delta^{\rm strain}_{\rm so}=2\hbar v_{\rm F}k^{\rm strain}_{\rm so}\approx 0.4$ meV, corresponding
to  $k^{\rm strain}_{\rm so}\approx 0.4\times 10^6$ m$^{-1}$ for $v_{\rm F}\approx 0.5\times 10^6$ m/sec. \cite{McEuen.2008}
For 
$d$ of the order of $\mu$m,
 $k^{\rm (strain)}_{\rm so}  d$ can therefore be of order $1-5$.

In this article we have presented a short review of recent theoretical predictions, which may bring new functionality to nanoscale devices through the electronic spin degree of freedom. The ``twisting" of the electronic spin induced by an SO interaction that is geometrically localized to a weak link between bulk electrodes, can be viewed as a splitting of electronic waves in spin space --- a phenomenon we call Rashba spin-splitting. 
A common feature of the investigated Rashba spin-splitting devices is the possibility to tune the electronic transport through an SO-active weak link mechanically and possibly also electrostatically by ``spin gating". \cite{Shekhter.2016}  We have shown that this is possible both for normal and superconducting electron transport. Nevertheless, more research has to be done in order to develop a complete theory of Rashba gating of normal and superconducting weak links. Here we would like to mention a few possible directions for future work.

{\em Role of  the  ``spin quantization axis" in the leads.}
The electron spin projection on an arbitrary chosen axis 
can take the two possible values $\pm \hbar/2$, meaning that these are the eigenvalues of a certain operator acting on the spin wave-function. The corresponding eigenfunctions span the full Hilbert space. When the electronic spin is decoupled from other degrees of freedom and external fields, this spin operator commutes with the Hamiltonian and therefore the same eigenfunctions are also eigenfunctions of the Hamiltonian and and thus represent  stationary spin states for any choice of the  spin quantization axis. This is no longer the situation  if the spin is coupled to an external magnetic field, in which case the eigenstates correspond to a spin quantization axis that is parallel to the magnetic field. It is also not the case for an SO-active material where the spin eigenstates correspond to wave-vector dependent directions of the spin quantization axis. When an SO-active weak link connects two SO-inactive electrodes, the spin quantization axis will in general point in different directions in different parts of the device. Which spin state the electrons occupy in the source electrode is therefore important. It follows that the choice of spin quantization axis in the leads, which can be accomplished by applying a weak external magnetic field (see the discussion in Sec. \ref{spin-selective}), is another tool for spin-controlled electron transport phenomena that needs to be fully investigated.

{\em Spin-vibron coupling in nano-electromechanical weak links.}
The sensitivity of spin-controlled transport through an SO-active weak link to a mechanical deformation of the link, which has been
demonstrated in this review, leads to the question of how transport is affected by the coupling between the spin and the mechanical vibrations. Such a coupling can be strong enough to cause spin-acoustic functionality in SO-active nano-electromechanical devices, which deserves to be investigated.

{\em Singlet-to-triplet spin conversion in spin-orbit active superconducting weak links.}
The possibility of an SO-induced ``spin redesign" of Cooper pairs passing through a Rashba weak link, which was demonstrated in Sec.~\ref{Josephson}, raises the question of what kind of links can be established between two superconductors based on pairs of electrons in different spin states. The spin polarization of Cooper pairs that may result from their propagation through an SO-active Rashba spin-splitter allows for a gradual change of their spin state and hence for a transformation between a spin-singlet Cooper pair and a spin-triplet pair. The complete theory of the above conversion can be connected to the interesting problem of proximity-induced spin polarization of superconductors.

{\em Role of the Coulomb interaction in spin-gated devices.}
Charge and spin are two fundamental properties of electrons and we have shown that, due to the SO interaction, spin as well as charge couples to an electric field. Therefore, the question of how electron transport through a weak link is affected by the interplay between the Coulomb blockade of  tunneling processes and the phenomenon of spin splitting is an intriguing task for future research. For example, in the study of a superconducting SO-active weak link \cite{Shekhter.2016b} (see also Sec. \ref{Josephson}), the Coulomb blockade phenomenon was used to simplify the process of spin polarization of a Cooper pair by decomposing it into a sequence of spin twists of single electrons. What the result of lifting the Coulomb blockade will be, is an important question for future research.

To conclude we emphasize that the study of spin-controlled transport through SO active weak links is only in its infancy. We believe that the early progress, some  of it   reviewed here, has laid a solid foundation for a wealth of future experimental and theoretical achievements.

\begin{acknowledgments}
This work was partially supported by the Swedish Research
Council (VR), by the Israel Science Foundation
(ISF) and by the infrastructure program of Israel
Ministry of Science and Technology under contract
3-11173.
 \end{acknowledgments}
 
 \appendix

\section{Tunneling elements}
\label{app.tun}

In a simplified model, the weak links between the electrodes  are pictured as straight segments connected at contorted bends. Considering the bent regions as scattering centers,  this Appendix outlines the derivation of the tunneling elements representing the weak links. 
The probability amplitude 
for tunneling from one lead to the other is constructed  from products of propagators along the straight segments and a  transfer amplitude across the contorted parts.

The propagator of the electron along a straight segment in the presence of spin-orbit interactions and a Zeeman field is  found in Sec. \ref{seg}. 
Confining the discussion to a bent wire that couples two electrodes,  the effective tunneling elements in-between them are derived in Sec. \ref{bend}. The effect of the mechanical degrees of freedom on the tunneling elements is introduced in Sec. \ref{pol}: the vibrations' dynamics is incorporated into the effective tunneling. \cite{Glazman}

\subsection{Propagation along a straight segment}
\label{seg}

The electron's propagation along a straight segment is described by the Green's function corresponding to the Hamiltonian  there. Assuming that the motion  is ballistic, the spatial part of the wave function is taken as a plane wave. The propagator from point 
${\bf r}'$ to point ${\bf r}$
is then
\begin{align}
G({\bf r}-{\bf r}';E)=\int dk e^{ik|{\bf r}-{\bf r}'|}[E+i0^{+}_{}-{\cal H}({\bf k})]_{}^{-1}\ ,
\label{G}
\end{align}
where $E$ is the electron's energy, and the vector ${\bf k}$, whose length is $k$, lies  in the direction of the segment connecting 
connecting ${\bf r}'$ with ${\bf r}$.
When the SO interaction is the one given in Eq.~(\ref{HAM}), the Hamiltonian is (adopting units in which $\hbar =1$)
\begin{align} 
{\cal H}({\bf k})=\frac{k^{2}}{2m^{\ast}}+\frac{k^{}_{\rm so}}{m^{\ast}}\sig\cdot({\bf k}\times\hat{\bf n})
\ .
\end{align}
Inserting this expression into Eq. (\ref{G}) and carrying out the integration over the length $k$, one obtains the propagator as given in 
Ref.~\onlinecite{Shahbazyan} (see also Ref.~\onlinecite{Ora.and.Amnon}).

Here we extend that calculation in two directions. First, we allow for other forms of the SO interaction. For instance, 
SO coupling may  be induced by strains, as happens in  carbon nanotubes,  where 
 the spin dynamics is described by the effective interaction  \cite{strain1,strain2}
\begin{align}
{\cal H}^{\rm strain}_{\rm so}=
v^{}_{\rm F}k^{\rm strain}_{\rm so}\hat{\bf k}\cdot\sig\ .
\label{strain}
\end{align}  
Here, $v_{\rm F}$ is the Fermi velocity [see also Eq.~(\ref{HAMstrain})]; 
 for $v_{\rm F}\approx 
0.5 \times 10^{6}$ m/sec, one finds $k_{\rm so}^{\rm strain}\approx 0.4\times 10^{6}$ m$^{-1}$. \cite{Kuemmeth}
Second, one may wish to find the propagator in the presence of a magnetic field ${\bf B}$. The orbital effect of this  field on the motion along a one-dimensional wire can be accounted for by assigning an Aharonov-Bohm phase factor to the propagator,  the phase being the magnetic flux (in units of the flux quantum)  accumulated from the field  upon moving along the segment. (Naturally, this phase factor depends on the choice of the coordinate origin; the  physical quantities, however,   include only  the total Aharonov-Bohm flux through closed loops. \cite{Aharonov.Bohm}) The  magnetic field is coupled also to the spin, adding to the the Hamiltonian ${\cal H}({\bf k})$ the Zeeman interaction, 
$\mu_{\rm B}{\bf B}\cdot\sig$. 
It follows that the generic form of the (ballistic) Hamiltonian on the straight segment is
\begin{align}
{\cal H}({\bf k})=\frac{k^{2}}{2m^{\ast}}+{\bf Q}({\bf k})\cdot\sig\ .
\label{gen}
\end{align}
For example, for the  SO interaction  Eq.~(\ref{HAM}), 
\begin{align}
{\bf Q}({\bf k})=\frac{k^{}_{\rm so}}{m^{\ast}}({\bf k}\times\hat{\bf n})+\mu^{}_{\rm B}{\bf B}\ .
\label{vecQ}
\end{align}
Note that ${\bf Q}({\bf k})$ combines together the Zeeman magnetic field, and the effective magnetic field representing the SO interaction.

The Hamiltonian (\ref{gen}) is easily diagonalized:
the eigenvalues are 
\begin{align}
\epsilon^{}_{\pm}({\bf k})=\frac{k^{2}}{2m^{\ast}}\pm Q({\bf k})\ ,
\end{align}
and the projection operators into each of the corresponding  subspaces 
are
\begin{align}
\Pi^{}_{\pm}({\bf k})=\frac{1\pm\hat{\bf Q}({\bf k})\cdot\sig}{2}\ ,
\end{align}
where $\hat{\bf Q}({\bf k})$ is a unit vector in the direction of the vector ${\bf Q}({\bf k})$, whose length is $Q({\bf k})$.
Using the diagonalized form in Eq.~(\ref{G}) yields
\begin{align}
G({\bf r}&-{\bf r}';E)=\int dk e^{ik|{\bf r}-{\bf r}'|}\nonumber\\
&\times\Big (\frac{\Pi^{}_{+}({\bf k})}{E+i0^{+}_{}-\epsilon^{}_{+}({\bf k})}+
\frac{\Pi^{}_{-}({\bf k})}{E+i0^{+}_{}-\epsilon^{}_{-}({\bf k})}\Big )\ .
\label{G1}
\end{align}
The poles  of the integrand in Eq.~(\ref{G1}) are given by the relation
\begin{align}
k^{2}_{\pm}=k^{2}_{0}\mp 2m^{\ast} Q(k)+i0^{+}_{}\ ,
\end{align}
where $k^{2}_{0}=2m^{\ast}E$. Hence, 
\begin{align}
G({\bf r}-{\bf r}';E)&=i\pi\Big (\frac{m^{\ast}e^{ik^{}_{+}|{\bf r}-{\bf r}'|}}{k^{}_{+}} 
\Pi^{}_{+}(k^{}_{+})\nonumber\\
&+
\frac{m^{\ast}e^{ik^{}_{-}|{\bf r}-{\bf r}'|}}{k^{}_{-}} 
\Pi^{}_{+}(k^{}_{-})
\Big )\ .
\label{G2}
\end{align}
(Note that the angles of the vector ${\bf k}$ are not changed along the straight segment, and therefore the integration is carried out over the magnitude, $k$.) The energy $E$ corresponds to the Fermi energy in the leads; assuming that it is much larger than the energy scales of the SO interaction and the magnetic field, one may use the approximation
\begin{align}
k^{}_{\pm}\approx k^{}_{0}\mp\frac{m^{\ast}}{k^{}_{0}}Q(k^{}_{0})\ ,
\end{align}
to obtain
\begin{widetext}
\begin{align}
G({\bf r}-{\bf r}';E)&=G^{}_{0}({\bf r}-{\bf r}';E)
\Big (e^{-i\frac{m^{\ast}Q^{}_{0}|{\bf r}-{\bf r}'|}{k^{}_{0}}}\Pi^{}_{+}(k^{}_{0})+e^{i\frac{m^{\ast}Q^{}_{0}|{\bf r}-{\bf r}'|}{k^{}_{0}}}\Pi^{}_{-}(k^{}_{0})\Big )\nonumber\\
&=G^{}_{0}({\bf r}-{\bf r}';E)\Big (\cos\Big [\frac{m^{\ast}Q^{}_{0}|{\bf r}-{\bf r}'|}{k^{}_{0}}\Big ]-i\sin\Big [\frac{m^{\ast}Q^{}_{0}|{\bf r}-{\bf r}'|}{k^{}_{0}}\Big ]\sig\cdot\hat{\bf Q}^{}_{0}\Big )\ .
\label{G3}
\end{align}
\end{widetext}
Here, 
\begin{align}
G^{}_{0}({\bf r}-{\bf r}';E)=i\pi (m^{\ast}_{}/k^{}_{0})\exp[ik^{}_{0}|{\bf r}-{\bf r}'|]\ ,
\end{align}
is the propagator on the segment in the absence of the SO interaction and the magnetic field, and
\begin{align}
{\bf Q}^{}_{0}={\bf Q}(k^{}_{0})\ , 
\label{Q0}
\end{align}
where the angles of the  vector ${\bf k}$ are those of the straight segment.  The spin dynamic, caused by the spin-orbit interaction and the Zeeman field, is contained in the second factor of Eq.~(\ref{G3}).

\subsection{Weak links  with a bend}
\label{bend}
Figure \ref{Fig7}(b) illustrates the  model system used in the calculations. The  weak link between two electrodes, taken to lie in the $XY$ plane,  is 
replaced by two straight one-dimensional wires, ${\bf R}_{L}$ and ${\bf R}_{R}$, of 
equal length $d/2$, connected by a bent. The angles between these wires and the $\hat{\bf x} $ axis 
are $\theta$ and $-\theta$, respectively. This means that the direction of the vector ${\bf k}$ [see, e.g., Eq.~(\ref{vecQ})]
of the left wire is $\{\cos\theta,\sin\theta,0\}$ and  that of the right wire is
$\{\cos\theta,-\sin\theta,0\}$.  These unit vectors determine the corresponding vectors ${\bf Q}_{0}$, Eq.~(\ref{Q0}).
For this configuration, 
the tunneling amplitude, a 2$\times$2 matrix in spin space, is
\begin{align}
&W=W^{}_{0}{\cal W}\ ,
\label{et}
\end{align}
with
\begin{align}
W^{}_{0}=
G^{}_{0}(|R^{}_{L}|;E)
{\cal T}
G^{}_{0}(|R^{}_{R}|;E)
\ ,
\label{W0}
\end{align}
where
${\cal T}$ is
the transfer matrix through the bent in the wire.
This scalar amplitude comprises all the characteristics of the tunneling element that are independent of the spin dynamics.  The latter  is embedded in the matrix ${\cal W}$, 
\begin{align}
{\cal W}&=\exp[-i\psi^{}_{L}]\times\exp[-i\psi^{}_{R}]\ ,
\label{calw}
\end{align}
where
\begin{align}
\psi^{}_{L(R)}=\frac{m^{\ast}_{}Q^{}_{0L(R)}d}{2k^{}_{0}}\sig\cdot\hat{\bf Q}_{0L(R)}\ .
\label{psi}
\end{align}
The unitary matrix ${\cal W}$
performs two consecutive spin rotations of the spins,
around the 
axes $\hat{\bf Q}_{0L}$ and $\hat{\bf Q}_{0R}$.
For the SO interaction given in Eq.~(\ref{HAM}), and in the absence of the Zeeman field, one finds that $Q_{0L}=Q_{0R}=k^{}_{0}k^{}_{\rm so}/m^{\ast}$, and ${\bf Q}_{0L(R)}=(d/2)\hat{\bf R}_{L(R)}\times\hat{\bf n}$.
Equation 
(\ref{et}) 
is derived to lowest possible order in the tunneling; the explicit dependence of $W_{0}$ 
on the momenta
is omitted for brevity.

\subsection{Vibrational degrees of freedom}
\label{pol}

Coupling the charge carriers with the  mechanical vibrations of the suspended  nanowire forming the junction adds an interesting aspect to the tunneling elements. For example, it was shown that this coupling can render  the conductance  through a single-channel wire to be affected by a constant magnetic field. The bending vibrations 
modify
geometrically the spatial region where an orbital magnetic field
is present, leading to a finite Aharonov-Bohm effect,  \cite{Glazman} which in turn gives rise to a magnetic-field dependence of the transmission. 
Likewise, the effect of the SO interaction can be modified by the effective area covered by the vibrating wire.  \cite{Shekhter.2013.2014}

Consider for instance the setup depicted in  Fig.~ \ref{Fig7}(b).
Within  this plausible geometry, $y_{L}=y_{R}=(d/2) \sin\theta$ and $x_{L}=x_{R}=(d/2) \cos\theta$, where $\theta$ is the instantaneous  bending angle. (An alternative geometry, with $x_{L}=x_{R}=d/2$ and $y_{L}=-y_{R}=(d/2){\rm tan}\theta$,  gives similar results.)
In order to mimic
the bending vibrations of the wire we assume that once the
wire is bent by the (equilibrium) angle $\theta_{0}$, 
then the distance
along $x$
between the two supporting leads is fixed, while the
bending point 
vibrates along
$y$.
As a result, ${\rm tan}\theta =2y/[d\cos\theta]$, implying that $\Delta\theta =(2/[d\cos\theta])
\cos^{2}\theta_{0}\Delta y$. (Here $d\cos\theta_{0}$ is the wire projection on the $\hat{\bf x}$ direction.])
It
follows that
\begin{align}
\theta =\theta^{}_{0}+\Delta\theta =\theta^{}_{0}+(a^{}_{0}\cos\theta^{}_{0}/d)(b+b^{\dagger}_{})\ ,
\label{tet}
\end{align}
where $a_{0}$
is the amplitude of the zero-point oscillations and
$b$ ($b^{\dagger})$
 is the annihialtion  (creation) operator of the vibrations.
Their free Hamiltonian is described by the Einstein model, 
\begin{align}
{\cal H}_{\rm vib}=\omega^{}_{0}b^{\dagger}b\ .
\label{ein}
\end{align}
Details of the derivation of the current through a vibrating wire are given in Sec. \ref{vib}.

\section{Spin-resolved currents}
\label{app.src}

This Appendix is divided  into two parts.  The spin-resolved currents
through the weak link discussed in Secs. \ref{seg} and \ref{bend} are derived in Sec. \ref{stat}; that part ignores   the effect of the mechanical vibrations.
For the sake of completeness, we allow for the possibilities that the junction couples  two superconducting electrodes, 
a superconducting and a normal one, or two normal-state electrodes. In all these cases we assume that each electrode is described by a free electron gas, augmented (in the case of a superconducting lead) by the BCS Hamiltonian. The currents through a vibrating nanowire are considered in Sec. \ref{vib}. For simplicity the discussion there is confined to a junction connecting two normal electrodes. 

\subsection{Spin-resolved current  through static weak links}
\label{stat}

We consider
the simplified,  though realistic, 
model, in which two electrodes are connected  via a spin-dependent tunnel Hamiltonian, 
\begin{align}
{\cal H}^{}_{T}=\sum_{{\bf k},{\bf p}}\sum_{\sigma,\sigma '}(c^{\dagger}_{{\bf p}\sigma '}[W^{}_{{\bf p},{\bf k}}]^{}_{\sigma',\sigma}c^{}_{{\bf k}\sigma}+{\rm H.c.})\ .
\label{HT}
\end{align}
Here, 
\begin{align}
[W^{}_{{\bf p},{\bf k}}]^{}_{\sigma,\sigma '}=([W^{}_{-{\bf p},-{
\bf k}}]^{}_{-\sigma,-\sigma '})^{\ast}
\label{relW}
\end{align}
are elements of a matrix in spin space, which obey time-reversal symmetry.  \cite{Anderson} (In the presence of a Zeeman interaction the sign of the magnetic field in the matrix element on the right-hand side is reversed.)  The relation   (\ref{relW}) adds to the one  imposed by the hermiticity of the Hamiltonian, 
\begin{align}
[W^{}_{{\bf p},{\bf k}}]^{}_{\sigma,\sigma '}=([W^{}_{{\bf k},
{\bf p}}]^{}_{\sigma',\sigma })^{\ast}\ .
\label{relW1}
\end{align}
The operator $c^{\dagger}_{{\bf k}({\bf p})\sigma} $ creates an electron in the left (right) electrode, with momentum ${\bf k}({\bf p})$
and a spin index $\sigma$, which denotes the eigenvalue of the spin projection along an arbitrary axis.
The construction of the matrix elements is detailed  in Secs. \ref{seg} and \ref{bend}.

As mentioned, the electrodes are considered as BCS superconductors, 
\begin{align}
{\cal H}^{}_{L(R)} &= \sum_{{\bf k}({\bf p})} \xi^{}_{{\bf k}({\bf p})} c^\dag_{{\bf k}({\bf p})\sigma}c^{}_{{\bf k}({\bf p})\sigma} \nonumber \\
&+\Big ( \Delta_{L(R)} \sum_{{\bf k}({\bf p})} c^\dag_{{\bf k}({\bf p})\uparrow}c^\dag_{{-\bf k}(-{\bf p}),\downarrow} + {\rm H.c.} \Big )\ ,
\label{HLR}
\end{align}
where
$\xi_{{\bf k}({\bf p})} = \epsilon_{{\bf k}({\bf p})} - \mu$
is the quasi-particle energy in the left (right) bulk superconducting lead, and $\mu$ is the common chemical potential.    The superconductor  order parameter $\Delta_{L(R)}$ is given by the self-consistency relation 
\begin{align}
\Delta^{}_{L(R)}=V^{}_{\rm BCS}\sum_{{\bf k}({\bf p})}\langle c^{}_{-{\bf k}(-{\bf p})\down}c^{}_{{\bf k}({\bf p})\up}\rangle\ ,
\label{delta}
\end{align}
where $V_{\rm BCS} $ denotes the attractive interaction among the electrons. The total Hamiltonian of the junction is thus
\begin{align}
{\cal H}={\cal H}^{}_{L}+{\cal H}^{}_{R}+{\cal H}^{}_{T}\ .
\label{fullH}
\end{align}
Additional comments on the calculation of the current in-between two superconducting leads are given below. \cite{Cuevas}

The spin-resolved 
particle current emerging from the 
left electrode, $I_{L,\sigma}$,  is found by calculating the time 
evolution of the number operator of electrons with spin projection $\sigma$, $\dot{N}_{L,\sigma}$, 
\begin{align}
&-I^{}_{L,\sigma}\equiv \langle\dot{N}^{}_{L,\sigma  }\rangle =\frac{d}{dt}\Big \langle \sum_{\bf k}c^{\dagger}_{{\bf k}\sigma} c^{}_{{\bf k}\sigma}\Big \rangle\nonumber\\
&= 2{\rm Im} \sum_{{\bf k},{\bf p}}\sum_{\sigma'}
\Big \langle [W^{}_{{\bf p},{\bf k}}]^{}_{\sigma',\sigma}c^{\dagger}_{{\bf p}\sigma'}c^{}_{{\bf k}\sigma}
\Big \rangle\ ,
\label{I1}
\end{align}
where we have used the relation (\ref{relW1}) and the self-consistency requirement 
(\ref{delta}).

The angular brackets in Eq.~(\ref{I1}) denote the quantum-thermal average,  which we calculate  within second-order perturbation theory in the tunneling Hamiltonian ${\cal H}^{}_{T}$, Eq.~(\ref{HT}), 
\begin{align}
&I^{}_{L,\sigma}=
2\,{\rm Re}\sum_{{\bf k},{\bf  p}}\sum_{\sigma '}\nonumber\\
&\times\int_{-\infty}^{t}dt' \Big\langle  \Big
[[W^{}_{ {\bf p},{\bf k}}]^{}_{\sigma ',\sigma }c^{\dagger}_{{\bf p}\sigma '}(t)c^{}_{{\bf k}\sigma}(t),{\cal H}^{}_{T}(t')
\Big ]
  \Big\rangle\ .
  \label{I2}
\end{align}
The time-dependence of the operators should be handled with care. When both electrodes are superconducting the  difference  between the phases of the  two order parameters evolves in time according to the Josephson relation, leading to an ac current. This is not taken into account in the second-order perturbation calculation presented below, and therefore
when the junction couples two superconducting leads, 
our treatment is valid only for the equilibrium situation, where no bias is applied across the junction. In that case the quasi-particle current vanishes.  However, the comparison between the amplitude of the current in the normal state of the junction,
i.e., the transmission of the junction, and that of the Josephson current, is of great interest since the SO interaction modifies them differently.
For this reason both currents are kept. In the standard perturbation calculation carried out here, the normal-state transmission is derived from that of the quasi-particles. 
Accordingly, the particle current is separated into two parts,
\begin{align}
I^{}_{L,\sigma}=I^{S}_{L,\sigma}+I^{N}_{L,\sigma}\ ,
\end{align}
with
\begin{widetext}
\begin{align}
I^{S}_{L,\sigma}=&
2\, {\rm Re}
\sum_{{\bf k },{\bf p}}\sum_{\sigma '}\int_{-\infty}^{t}dt'
|[ W^{}_{{\bf p},{\bf k}}]^{}_{\sigma ',\sigma }|^{2}
\Big \langle c^{\dagger}_{{\bf p}\sigma '}(t)c^{}_{{\bf k}\sigma}(t)c^{\dagger}_{-{\bf p}-\sigma '}(t')c^{}_{-{\bf k}-\sigma}(t')
-
c^{\dagger}_{-{\bf p}-\sigma' }(t')c^{}_{-{\bf k}-\sigma}(t')
c^{\dagger}_{{\bf p}\sigma '}(t)c^{}_{{\bf k}\sigma}(t)\Big  \rangle\ ,
\label{IS}
\end{align}
and
\begin{align}
I^{N}_{L,\sigma}=&2\,{\rm Re}
\sum_{{\bf k },{\bf p}}\sum_{\sigma '}\int_{-\infty}^{t}dt'
|[W^{}_{{\bf p},{\bf k}}]^{}_{\sigma ',\sigma }
|^{2}_{}
\Big \langle c^{\dagger}_{{\bf p}\sigma '}(t)c^{}_{{\bf k}\sigma}(t)
c^{\dagger}_{{\bf k}\sigma}(t')c^{}_{{\bf p}\sigma '}(t')
-
c^{\dagger}_{{\bf k}\sigma}(t')c^{}_{{\bf p}\sigma '}(t')c^{\dagger}_{{\bf p}\sigma '}(t)c^{}_{{\bf k}\sigma}(t) \Big \rangle\ .
\label{IN}
\end{align}

The quantum-thermal average is found by introducing the Green's functions of the bulk (left) lead,
\begin{align}
\hat{G}^{+-}({\bf k};t,t')=i
\left [
\begin{array}{cc}\langle c^{\dagger}_{{\bf k}\up}(t')c^{}_{{\bf k}\up}(t)\rangle &\ \
\langle c^{}_{-{\bf k}\down}(t')c^{}_{{\bf k}\up}(t) \rangle\\ \langle c^{\dagger}_{{\bf k}\up}(t')c^{\dagger}_{-{\bf k}\down}(t)\rangle &\ \ \ \
\langle c^{}_{-{\bf k}\down}(t')c^{\dagger}_{-{\bf k}\down}(t)\rangle \end{array}\right ]\ ,\ \ \  \hat{G}^{-+}({\bf k};t,t')=
[
\hat{G}^{+-}({\bf k};t,t')]^{\ast}_{}\ ,
\label{pm}
\end{align}
and their Fourier transforms,
\begin{align}
&\hat{G}^{+-}_{}({\bf k},\omega)= i[1-f^{}_{L}(\omega)]
\left [\begin{array}
{cc}A^{}_{\bf k}(\omega)
&\ \ \ \ \ \ e^{i\varphi^{}_{L}}B^{}_{\bf k}(\omega )
\\
e^{-i\varphi^{}_{L}}B^{}_{\bf k}(\omega )
&\ \ \ \ \ \
A^{}_{\bf k}(-\omega)
\end{array}\right ]\ ,\ \ \ \ \hat{G}^{-+}_{}({\bf k},\omega)=-\frac{f^{}_{L}(\omega)}{1-f^{}_{L}(\omega)}\hat{G}^{+-}_{}({\bf k},\omega)\ .
\label{gfour}
\end{align}
Analogous expressions pertain for the right lead, with ${\bf k}$ replaced by ${\bf p}$, and $L$ by $R$.
The superconducting gap function of the left BCS lead is $\Delta^{}_{L}=|\Delta^{}_{L}|\exp[i\varphi^{}_{L}]$,
and the coherence factors there are
$u^{}_{k}= |u^{}_{k}|\exp[-i\varphi^{}_{L}/2]$ and $v^{}_{k}= |v^{}_{k}|\exp[i\varphi^{}_{L}/2]$, with $|u^{}_{k}|^{2}=1-|v^{}_{k}|^{2}=
(1+\xi^{}_{k}/E^{}_{k}
)/2$,   
and
$E_{k}=\sqrt{\xi^{2}_{k}+|\Delta^{}_{L}|^{2}}$.
The spectral functions in Eq.~(\ref{gfour}) are
\begin{align}
A^{}_{\bf k}(\omega)&=2\pi
[|u^{}_{k}|^{2}\delta (\omega +E^{}_{k})+|v^{}_{k}|^{2}\delta (\omega -E^{}_{k})]\ ,\nonumber\\
B^{}_{\bf k}(\omega )&=-2\pi |u^{}_{k}v^{}_{k}|
[\delta (\omega +E^{}_{k})-\delta (\omega -E^{}_{k})]\ ,
\label{AB}
\end{align}
and
$f^{}_{L}(\omega)$  is the Fermi function of the quasi particles in the  left lead.

Inserting the relations (\ref{pm}) and (\ref{gfour}) into Eq.~(\ref{IS}) gives the equilibrium Josephson current, 
\begin{align}
I^{S}_{L,\sigma}&=\sin(\varphi^{}_{L}-\varphi^{}_{R})
\sum_{{\bf k},{\bf p}}
{\cal P}\int\frac{d\omega d\omega '}{2\pi^{2}}
\frac{f(\omega )-f(\omega ')}{\omega -\omega '}B^{}_{\bf p}(\omega)B^{}_{\bf k}(\omega ')
\{|[W^{}_{{\bf p},{\bf k}}]^{}_{\sigma,\sigma}|^{2}_{}
-|[W^{}_{{\bf p},{\bf  k}}]^{}_{\overline{\sigma},\sigma}|^{2}_{}
\}
\ ,
\label{JOS}
\end{align}
where we have used the symmetry $B_{\bf k}(\omega)=B^{}_{-{\bf k}}(\omega)=-B^{}_{\bf k}(-\omega)$ and  $f^{}_{L}(\omega)=f^{}_{R}(\omega)\equiv f(\omega)=(\exp[\beta\omega]+1)^{-1}$ ($\beta$ is the inverse temperature), since  as mentioned, the supercurrent is calculated at equilibrium; ${\cal P}$ denotes the principal part, and 
$\overline{\sigma}$ is the spin direction opposite to $\sigma$.
The transmission of the junction in the normal state is found by inserting the relations (\ref{pm}) and (\ref{gfour}) into Eq.~(\ref{IN})
for the normal part of the spin-resolved current, 
\begin{align}
I^{N}_{L,\sigma}&=\sum_{{\bf k},{\bf p}}
\int\frac{d\omega}{2\pi}[f^{}_{L}(\omega)-f^{}_{R}(\omega )]
\Big (|[W^{}_{{\bf p},{\bf k}}]^{}_{\sigma,\sigma}|^{2}+
|[W^{}_{{\bf p},{\bf k}}]^{}_{\overline{\sigma},\sigma}|^{2}
\Big )
A^{}_{\bf k}(\omega)A^{}_{\bf p}(\omega)\ .
\label{IL2}
\end{align}
Here we have used the symmetry $A_{\bf k}(\omega)=A_{-{\bf k}}(\omega )$.
As mentioned,  the quasi-particle current $I_{L}^{N}$, Eq.~(\ref{IL2}), vanishes for the unbiased junction for which $f_{L}(\omega )=f_{R}(\omega)$.

A comparison of  the two expressions, Eq.~(\ref{JOS}) and Eq.~(\ref{IL2}), 
reveals the different ways by which the total effective  magnetic field [the Zeeman field and the effective magnetic field due to the SO interaction, see Eq.~(\ref{gen})]
affects the Josephson current and  the  particle current in the normal state.
One notes that the diagonal (in spin space) elements of the tunneling matrix appear in these two expressions with the same sign, as opposed to the off-diagonal ones.
This implies 
that there is a significant difference between the  effect of the component of an effective magnetic field normal to the junction plane, and an effective magnetic field   in the junction's plane.

\subsection{Spin-resolved currents through vibrating nanowires}
\label{vib}

When the wire connecting the two electrodes [see Fig.~\ref{Fig7}(b)] is vibrating, the tunneling amplitude is a dynamical variable (see Sec. \ref{pol}), and hence depends on time. This modifies the calculation of the current. For convenience, the discussion is confined to the case where the weak link connects two normal electrodes. In second-order perturbation theory, the starting point is still Eq.~(\ref{I2}), but one has to include in the calculation the time dependence of the hopping amplitude. As a result, Eq.~(\ref{I2}) is modified,  
\begin{align}
I^{}_{L,\sigma}&=2 {\rm Re}\sum_{{\bf k},{\bf p}}\sum_{\sigma '}\int_{-\infty}^{t}dt'\Big \langle 
 \Big
[[W^{}_{ {\bf p},{\bf k}}]^{}_{\sigma ',\sigma }(t)c^{\dagger}_{{\bf p}\sigma '}(t)c^{}_{{\bf k}\sigma}(t),{\cal H}^{}_{T}(t')
\Big ]
  \Big\rangle\nonumber\\
 &=2 {\rm Re}\sum_{{\bf k},{\bf p}}\sum_{\sigma '}\int_{-\infty}^{t}dt'\Big (\langle [W^{}_{{\bf p},{\bf k}}]^{}_{\sigma ',\sigma}(t)[W^{}_{{\bf k},{\bf p}}]^{}_{\sigma ,\sigma'}(t')\rangle \langle c^{}_{{\bf k}\sigma}(t)c^{\dagger}_{{\bf k}\sigma}(t')\rangle\langle c^{\dagger}_{{\bf p}\sigma '}(t)c^{}_{{\bf p}\sigma '}(t')\rangle\nonumber\\
 &-\langle [W^{}_{{\bf k},{\bf p}}]^{}_{\sigma ,\sigma'}(t')[W^{}_{{\bf p},{\bf k}}]^{}_{\sigma ',\sigma}(t)\rangle \langle c^{\dagger}_{{\bf k}\sigma}(t')c^{}_{{\bf k}\sigma}(t)\rangle\langle c^{}_{{\bf p}\sigma '}(t')c^{\dagger}_{{\bf p}\sigma '}(t)\rangle \Big )\ .
\label{IP}
\end{align}
The quantum thermal averages over the electronic operators 
are obtained as in Sec. \ref{stat}, using Eqs. 
(\ref{pm}) and (\ref{gfour}) (with $\Delta_{L}=\Delta_{R}=0$). 
The thermal average over the tunneling amplitudes is carried out with the Einstein Hamiltonian, Eq.~(\ref{ein}).
Using the notations introduced in Eqs. (\ref{W0}),  (\ref{calw}),  and (\ref{psi}), 
we find
\begin{align}
\langle [W^{}_{{\bf p},{\bf k}}]^{}_{\sigma ',\sigma}(t)[W^{}_{{\bf k},{\bf p}}]^{}_{\sigma ,\sigma'}(t')\rangle &=|W^{}_{0}|^{2}\langle  [e^{-i\psi^{}_{R}(t)}\times e^{-i\psi^{}_{L}(t)}]^{}_{\sigma',\sigma}
 [e^{-i\psi^{}_{L}(t')}\times e^{-i\psi^{}_{R}(t')}]^{}_{\sigma,\sigma'}\rangle \nonumber\\
 &=\sum_{n,n'}P(n)e^{i(n-n')\omega^{}_{0}(t-t')}|\langle n|[e^{-i\psi^{}_{R}}\times e^{-i\psi^{}_{L}}]^{}_{\sigma',\sigma}|n'\rangle |^{2}\ .
 \end{align}
Here $|n\rangle$ indexes  the eigenfunctions of the Einstein Hamiltonian and 
\begin{align}
P(n)=\frac{e^{-n\beta\omega^{}_{0}}}{{\rm Tr}e^{-\beta{\cal H}^{}_{\rm vib}}}=e^{-n\beta\omega^{}_{0}}
(1-e^{-\beta\omega^{}_{0}})\ ,
\label{P}
\end{align}
auch that
\begin{align}
\sum_{n=0}^{\infty}P(n)=1\ ,\ \ \sum_{n=0}^{\infty}P(n)n=\frac{1}{e^{\beta\omega^{}_{0}}-1}\equiv N^{}_{\rm B}(\omega^{}_{0})\ .
\end{align}
Inserting the expressions for the quantum thermal averages into Eq.~(\ref{IP}) yields
\begin{align}
I^{}_{L,\sigma}&=2|W^{}_{0}|^{2}\int \frac{d\omega d\omega '}{4\pi}{\cal N}_{L,\sigma} \sum_{\sigma '}{\cal N}^{}_{R,\sigma '}\sum_{n,n'}P(n)\nonumber\\
&\times\Big (f^{}_{L\sigma}(\omega )[1-f^{}_{R\sigma '}(\omega ')]\delta [\omega -\omega '+(n-n')\omega^{}_{0}]
|\langle n|[e^{-i\psi^{}_{R}}\times e^{-i\psi^{}_{L}}]^{}_{\sigma ',\sigma}|n'\rangle |^{2}\nonumber\\
&-f^{}_{R\sigma'}(\omega' )[1-f^{}_{L\sigma }(\omega )]\delta [\omega ' -\omega +(n-n')\omega^{}_{0}]
|\langle n|[e^{i\psi^{\dagger}_{L}}\times e^{i\psi^{\dagger}_{R}}]^{}_{\sigma ,\sigma'}|n'\rangle |^{2}\Big )\ .
\label{SSD}
\end{align}
\end{widetext}
Here, ${\cal N}_{L(R),\sigma}$ are the spin-resolved densities of states at the common chemical potential of the device, $\mu=(\mu_{L}+\mu_{R})/2$ [see Eqs. (\ref{mlr}) and (\ref{MLR}), and the discussion following the latter].
The reservoirs are represented by their respective electronic distributions  determined by the spin-dependent electrochemical potentials, 
\begin{align}
&f^{}_{L,\sigma}(\epsilon^{}_{k,\sigma})=[e^{\beta (\epsilon^{}_{k,\sigma}-\mu^{}_{L,\sigma})}+1]^{-1}\ ,\nonumber\\
&f^{}_{R,\sigma '}(\epsilon^{}_{p,\sigma '})=[e^{\beta (\epsilon^{}_{p,\sigma '}-\mu^{}_{R,\sigma '})}+1]^{-1}\ ,
\label{Fermi}
\end{align}
with $\beta^{-1}=k_{\rm B}T$.

As expected, the coupling with the vibrational modes of the wire introduces inelastic processes into the tunneling current, in which the charge carriers exchange energy with the mechanical degrees of freedom. 
Another point to notice it that the current is not spin-resolved unless the electrodes are polarized. This point is further discussed in the main text. It reflects the conclusion reached in Sec. \ref{stat}: the contributions to the normal-state particle current
coming from the diagonal elements of the tunneling amplitude and that of the off diagonal ones add together.


\begin{thebibliography}{99}
\bibitem{com2}
Note  the peculiar way of defining the transmission as given  in Eqs.  (\ref{Tt}) and (\ref{T}): it has dimensions of energy squared because it is not yet multiplied by the appropriate densities of states. 
\bibitem{com6}
Because of charge conservation one has  $\sum_{\sigma}I^{\rm spin}_{\sigma}=2I^{\rm spin}_{\up}=-2I^{\rm spin}_{\down}$.
\bibitem{note}
In order to keep the estimates simple, it is assumed that both the Fermi velocities, $v_{\rm F}$,  and the superconducting gaps of the two leads are the same.
\end{thebibliography}
\end{document}